\documentclass[11pt,letterpaper]{article}
\pdfoutput=1

\usepackage{graphicx,array}
\usepackage[dvipsnames,table,xcdraw]{xcolor}
\definecolor{darkblue}{rgb}{0,0.1,0.5}
\definecolor{darkgreen}{rgb}{0,0.5,0.2}
\definecolor{darkred}{RGB}{153,26,0}
\usepackage{ulem}
\usepackage{slashed}
\definecolor{seablue}{rgb}{0,0.2,0.6}
\usepackage{latexsym}
\usepackage{amssymb, amsmath}
\usepackage{slashed}
\usepackage{bm}        
\usepackage[numbers,sort&compress]{natbib}
\usepackage{bm,relsize}
\usepackage{slashed}
\usepackage{tcolorbox}
\definecolor{viola}{RGB}{134,41,198}
\usepackage{mathrsfs}
\usepackage{hyperref} 
\hypersetup{
    colorlinks=true,       
    linkcolor=darkblue,          
    citecolor=BrickRed,        
    filecolor=magenta,      
    urlcolor=MidnightBlue           
}
\usepackage[all]{hypcap} 
\usepackage{subcaption}

\usepackage{tikzfeynman}

\usepackage[font=small,labelfont=bf,labelsep=period]{caption}

\usepackage{stackengine}
\stackMath

\newcommand{\be}{\begin{equation}}
\newcommand{\ee}{\end{equation}}

\newcommand{\Mpl}{M_{\rm Pl}}
\newcommand{\im}{\mathrm{Im}}

\definecolor{lightergray}{rgb}{0.9,0.9,0.9}

\newcommand{\Scal}{\mathcal{S}}      

\begin{document}

\begin{flushright}

\end{flushright}
\vspace{-1cm}
\begin{center}
{\LARGE \bf 
Particle Production from Inhomogeneities: \\ the off-shell side of gravitational waves}\\
\bigskip\vspace{1cm}
{
\large Michele Redi, Andrea Tesi}
\\[7mm]
{\it \small
INFN Sezione di Firenze, Via G. Sansone 1, I-50019 Sesto Fiorentino, Italy\\
Department of Physics and Astronomy, University of Florence, Italy
}
\end{center}

\vspace{.2cm}

\centerline{\bf Abstract} 
\begin{quote}
We continue the study of particle production from gravitational inhomogeneities in the early Universe. Focusing on sources active on sub-horizon scales, we derive general expressions relating particle production to the unequal-time two-point function of the stress-energy tensor sourcing scalar, vector and tensor metric perturbations. The resulting particle yield probes the time-like support of this correlator, and in the tensor case the same object controls gravitational-wave emission when evaluated on the light-like support. This establishes a phenomenological link between dark matter production and gravitational wave signals, allowing the dark matter mass to be related to the amplitude of the stochastic gravitational wave background. Our results show that, on sub-horizon scales, particle production from inhomogeneous metric backgrounds practically reduces to gravitational scattering. This directly connects the formalism to gravitational freeze-in from the Standard Model thermal bath, while extending it to non-thermal and out-of-equilibrium sources. We apply the formalism to first-order phase transitions and discuss the associated production from scalar and tensor perturbations. The mechanism can efficiently populate gravitationally coupled dark sectors, especially when the perturbations are generated shortly after inflation.
\end{quote}

\vfill
\noindent\line(1,0){188}
{\scriptsize{ \\ E-mail:\texttt{   \href{mailto:michele.redi@fi.infn.it}{michele.redi@fi.infn.it}, \href{andrea.tesi@fi.infn.it}{andrea.tesi@fi.infn.it}}}}

\newpage

\tableofcontents

\section{Introduction}

The existence of primordial perturbations  is of fundamental importance for our understanding the universe as it sets up the initial conditions for the evolution. According to the prevailing picture, inflation produces small inhomogeneities that are adiabatic and with an almost flat power spectrum. These seeds grow under the pull of gravity eventually forming the complex structures that we observe  in the universe today. Large inhomogeneities could also be produced from local processes inside the horizon, such as phase transitions or even by inflation for short scale modes that are not observed in cosmology. Tensor inhomogeneities have in particular received a lot of interest recently as they lead to  potentially observable gravitational wave (GW) backgrounds.

In this work we explore effects related to the ubiquitous metric perturbations and in particular the generation of spectator particles in a perturbed Friedman-Lemaitre-Robertson-Walker (FLRW) metric. 
Particle production from inhomogeneities was first considered \cite{Zeldovich:1977vgo,Campos:1991ff,Cespedes:1989kh}, see \cite{Hu:2020luk} for review. More recently the subject was reconsidered in \cite{kopp,kopp2,Garani:2024isu}, in particular in connection with the possibility to produce dark matter (DM) . 
In \cite{Garani:2025qnm} we studied in great generality the production of a dark sector only coupled through gravity to the Standard Model (SM). The result turns out to be very simple. The production of particles from inhomogeneities is efficient when their mass is negligible, i.e. the momentum of the particles produced is larger than the mass. In this regime the abundance can be computed in closed form from the two-point function of the metric perturbations. In particular both scalar and tensor perturbations  can lead to significant abundances. The result becomes extremely simple for conformally coupled sectors where the abundance of particles is just proportional to the central charge $c_J$ of the CFT. This encompasses in one shot fermions, gauge fields and even strongly coupled CFTs in terms of single number $c_J$. 

One source of inhomogeneities is inflation that explains the structure of the universe at large scales observed in Cosmic-Microwave-background and in the galaxy distribution.
In \cite{Garani:2024isu} we studied DM production induced by scalar perturbations finding that it can be significant if the perturbations become large towards the end of inflation.
In this case particle production can be related to the power spectrum  at the end of inflation and it behaves differently for scalar and tensor perturbations. As an application one can construct scenarios where the DM is only gravitationally coupled to the SM. Because the nature of the effect is quantum, the abundance generated is small 
so that to reproduce the observed DM density the mass turns out to be very heavy, in the PeV range or larger.

Let us mention relation to other work. An extended literature exists on particle production due to the expansion of the universe during inflation or thereafter \cite{Kolb:2023ydq}.  Minimally coupled scalars or massive dark photons lead to light DM scenarios while particles such as fermions can only reproduce the cosmological abundance if they are very heavy. We emphasize that the nature of this effect is different from the production from inhomogeneities we discuss. The key of the previous mechanism is the explicit breaking of Weyl invariance that arises due to the mass or kinetic terms of elementary particles. For example a massless free fermion is described by a Weyl invariant action and as consequence is not produced -- up to anomalies -- in the homogeneous expanding universe. On the other hand a minimally coupled scalar is not Weyl invariant even in the massless limit, explaining the completely different behaviour. In the presence of inhomogeneities the situation is vastly different and all particles are produced. Depending on the details either inhomogeneities or the expansion might dominate the production but they are clearly independent. 

In this work we extend our previous study focusing on sources that produce inhomogeneities inside the horizon.  This includes cosmological phase transitions and preheating, topological defects  or thermal plasmas. In the sub-horizon regime it is simple and illuminating to write the formulas in terms of the energy momentum tensor that generates the metric perturbation. One finds, similarly to the production of GW, that the abundance is determined by the two-point function of the energy momentum tensor. While GW production depends on the two-point function in the light-like region of momenta, $q_\mu q^{\mu}=0$, particle production is determined by the time-like region, $q_\mu q^\mu>0$, corresponding for elementary particles to real production of particle pairs.

This analysis unifies and extends several previous computations.
A mechanism widely discussed in the literature is the production of a dark sector from the SM thermal plasma, also known as gravitational freeze-in \cite{Garny:2015sjg}. In \cite{Redi:2020ffc,Redi:2021ipn} we showed that this is also controlled by the central charge of the sector, hinting to a relation to particle production from inhomogeneities. 
Indeed in this work we show explicitly that gravitational freeze-in is equivalent to production from finite temperature scattering.  
Another important connection  that we uncover is with GW. For tensor perturbations the production of the dark sector depends on the same two-point function of the energy momentum tensor that determines a stochastic background of GW,  see \cite{Caprini:2018mtu} for a review. This implies a relation between the abundance of GW and of the one of dark sector particles that is of extreme interest. 
This  opens the way to numerical study in realistic scenarios such as first order phase transitions or preheating using the tools employed for stochastic GW background.

The organization of the paper is as follows. In section \ref{sec:review} we review the approach of \cite{Garani:2025qnm}, writing the explicit and general formula for the production of conformally coupled matter from gravitational inhomogeneities. In doing so, we highlight the role of the symmetries of the dark sector and clarify the question of finiteness of the result. While the formulas of section \ref{sec:review} are completely general, they can be further simplified when the inhomogeneities are generated by physics within the horizon. In this limit, that is discussed in section \ref{sec:micro}, we are able to connect the expressions of the gravitational inhomogeneities with the form of the stress-energy tensor of the system that is actually sourcing them. This discussion is inherently on sub-horizon scales, where we can explicitly relate the gravitational inhomogeneities (fluctuations) to the stress-energy tensor, solving the GW equations of motion.  
Using our formalism we give a novel derivation of gravitational freeze-in in section \ref{sec:GFI}.
In sections \ref{sec:PT} apply our results to first order phase transitions. We conclude in section \ref{sec:conclusions} with future directions. 
In the appendix we outline the generalization of our formulas to include the expansion of the universe during production.

\section{Production of conformally coupled sectors from Inhomogeneities}\label{sec:review}

To set the stage we start reviewing the main results of \cite{Redi:2021ipn} for gravitational production of conformally coupled matter from inhomogeneities. We consider an FLRW gravitational background with inhomogeneities described by,
\be\label{eq:metric}
ds^2= a^2(\tau)[ \eta_{\mu\nu}+ h_{\mu\nu}(\vec x\,, \tau)] dx^\mu dx^\nu\,,
\ee
and a spectator sector made of massless particles. Assuming that the sector is conformally coupled to the background it is possible 
to rescale the metric eliminating the scale factor from all computations\footnote{For production inside the horizon that we will mostly consider in this paper 
the assumption of conformality can be relaxed and one obtains completely analogous formulas with a different value of $c_J$.
We focus on massless particles because the production mechanism is strongly suppressed for massive particles requiring $q_\mu q^\mu\ge m^2$ where $q_\mu$ is the momentum of the perturbation.}. 
We can then easily compute the effective action as a function of background external field $h_{\mu\nu}(x)$ to second order in perturbations,
\begin{equation}
\Gamma_{\rm 1PI} =\frac {i} 8 \int \frac{d^4q}{(2\pi)^4} \frac{d^4q'}{(2\pi)^4}h^{\mu\nu}(-q) \langle  T_{\mu\nu}(q)   T_{\rho\sigma}(q')\rangle h^{\rho\sigma}(-q')\,,
\label{eq:G1PI}
\end{equation}
where the expectation value is time-ordered. Following Schwinger \cite{Schwinger:1951nm} twice the imaginary part of the effective action is just the probability of vacuum decay. 
As a consequence we can determine the inclusive number of particles produced by just computing  the two-point function of
the energy momentum tensor of the spectator sector in vacuum.
What is remarkable for conformally coupled sectors is that the two-point function is fixed up to an overall constant,
 (see for example \cite{Gubser:1997se})
\begin{equation}
\langle  T_{\mu\nu}(q)  T_{\rho \sigma }(q')\rangle =i(2 \pi)^4 \delta^4(q+q') \frac{c_J}{7680 \pi^2}\Pi_{\mu\nu\rho\sigma}(q)   \log(-q^2)\,,
\label{eq:2pointTT}
\end{equation}
where
\begin{equation}\label{projector}
\Pi_{\mu\nu\rho\sigma}(q) \equiv \left(2\pi_{\mu\nu} \pi_{\rho\sigma}-3\pi_{\mu\rho} \pi_{\nu\sigma}-3\pi_{\mu\sigma} \pi_{\nu\rho}\right)\,,\quad\quad \pi_{\mu\nu}\equiv\eta_{\mu\nu}q^2-q_\mu q_\nu\,.
\end{equation}
The structure of the two-point function is completely determined by conformal invariance up to the constant $c_J$ that is known as the central charge of the CFT.
In our normalization,  $c_0=4/3$, $c_{1/2}=4$ and $c_1=16$ respectively for conformally coupled scalars, Weyl fermions and massless gauge fields. 
Using  (\ref{eq:2pointTT})  we obtain, 
\begin{equation}
N_{\rm dec}\equiv 2{\rm Im}\Gamma_{\rm 1PI} =-\frac {c_J}{30720 \pi}  \int \frac{d^4q}{(2\pi)^4} \theta(q^2)  h_{\mu\nu}(q) \Pi^{\mu\nu\rho\sigma}(q) h_{\rho\sigma}(-q)\,.
\label{eq:Nparticles}
\end{equation}
In weakly coupled theories the energy momentum tensor is quadratic in the fields so that per each decay event two particles are produced. As a consequence 
the number of particles is just twice this result.
However, our derivation also applies to the class of interacting CFTs, since all that is needed is just $c_J$. In this case the imaginary part of the effective action counts the number of CFT shells being produced. 

The formula above is general. When the perturbation $h_{\mu\nu}$ has stochastic nature, we need to take the average $\langle h_{\mu\nu}(q)h_{\rho\sigma}(-q)\rangle$ in eq.~\eqref{eq:Nparticles}. 
In real time the result thus depends on the unequal time two-point function of the metric perturbation.
Note that for a static background $h_{\mu\nu}(q)\propto \delta(q_0)$ so  one can no particles are produced.
Time dependence can either arise due to the cosmological evolution of the gravitational field or due to the presence of sources. 
In \cite{Garani:2025qnm} scalar perturbations of the metric and GW from inflation were considered.  Inflation produces a power spectrum of  perturbations that are frozen once the exit the horizon during inflation and start to evolve when the modes re-enter the horizon. 
In this case the time dependence is exactly known  and depends only on the cosmological evolution. 
In particular while tensor perturbations are always small, scalar perturbations can be large at short scales, leading to significant production of particles.
In this note we will generalize this analysis to perturbations produced after inflation inside the horizon.

\paragraph{Weyl tensor and  Schwinger pair production}~\\
Not surprisingly given  Weyl invariance of the action the formula \eqref{eq:Nparticles} can also expressed in terms of the (linearized) Weyl tensor. One finds,
\be\label{eq:NparticlesWeyl}
N_{\rm dec}\equiv 2{\rm Im}\Gamma_{\rm 1PI} =\frac{c_J}{2560 \pi}  \int \frac{d^4q}{(2\pi)^4} \theta(q^2)  W_{\mu\nu\rho\sigma}(q) W^{\mu\nu\rho\sigma}(-q)\,.
\ee
This makes manifest the geometric nature of particle production from metric inhomogeneities.

This formulation allows to draw a close analogy with particle production from a varying electro-magnetic background.
At quadratic order in the external field $A_\mu$ the effective action can be written in terms of the two-point function of the electro-magnetic
current, $\langle J_\mu(q) J_\nu(-q)\rangle= (q^2\eta_{\mu\nu}-q_\mu q_\nu) \Pi(q^2)$. 
Considering for simplicity the massless case $\Pi(q^2)=e^2/(12\pi^2) q^2 \log (-q^2)$,  see also \cite{VicenteGarcia-Consuegra:2025lkh} for the massive case..
The imaginary part of the 1PI effective action at leading order in the external field $A_\mu$ is explicitly given by, 
\be
2\mathrm{Im}[\Gamma(A)] =\int \frac{d^4q}{(2\pi)^4} A_\mu (q) A_\nu(-q) \mathrm{Im}[\langle J^\mu(q)J^\nu(-q)\rangle] =  \frac{e^2}{12\pi} \int \frac{d^4q}{(2\pi)^4}\theta(q^2) (|\vec E(q)|^2- |\vec B(q)|^2)\,.
\ee
One can quickly check  $\vec E^2 >\vec B^2$ for $q^2>0$ so that the integrand is positive. This implies that the effect is electric.

One can show that very similar results hold for gravitational inhomogeneities, see \cite{Boasso:2024ryt} for a related discussion.
The Weyl tensor can be expressed in terms of its electric and magnetic components,
\be
W_{\mu\nu\rho\sigma}W^{\mu\nu\rho\sigma}=8(E_{ij}^2-B_{ij}^2)\,,\quad E_{ij}\equiv W_{0i0j}\,,\quad B_{ij}\equiv \frac12\epsilon_{ikl}W_{klj0}\,.
\ee
The electric and magnetic components are traceless, and the magnetic contribution is transverse thanks to the Bianchi identity. Exploiting these properties together with the Bianchi identities for the Weyl tensor, we can show that the square of the Weyl tensor is positive for electric configurations, by using $q_0 B_{ij} =\frac12(\epsilon_{ikl} q_k E_{lj}+\epsilon_{jkl} q_k E_{li})$, which is the analouge of the Maxwell's equation $q_0 \vec B = \vec q\times \vec E$. Indeed we can show
\be
W_{\mu\nu\rho\sigma}W^{\mu\nu\rho\sigma}=8\left(1-\frac{\vec q\,^2}{q_0^2}\right)E_{ij}^2 -\frac{3}{2 q_0^2} q_i E_{ik} q_j E_{jk}=\frac{4}{3}\vec{q}\,^4 \Theta^2 + 2(q_0^2-\vec q\,^2) \vec q\,^2 q_0^2 W_i^2    +\frac 1 2 (q_0^2-\vec q\,^2)^2 h_{ij}^2\,.
\ee
In the last step we have expressed the Weyl tensor in terms of scalar, $\Theta$, vector, $W$, and tensor, $h$, perturbations defined precisely below.
In the first equality the first term is positive on time-like supports (and there is a complete analogy with electromagnetism by replacing the Weyl tensor with the photon field strength), while the second term depends on the actual perturbation, as shown in the second equality. Scalar and tensor contributions are always positive (with the scalar one being purely electric). Vector contributions are positive on the time-like support. Notice that null-like tensor and vector perturbations do not produce particles.
\footnote{Explicitly for scalar perturbations, the electric and magnetic components of the Weyl tensor are $E_{ij}= \frac12(q_i q_j - q^2\delta_{ij}/3)\Theta$ and $B_{ij}=0$. For scalar perturbations we see that the presence of spatial gradient is essential to have particle production.}

\subsection{Production from Stochastic backgrounds}\label{sec:notation}

The master formula of eq.~\eqref{eq:Nparticles} can be specialised to various backgrounds. Perturbations of the metric, that can be classified into scalar, vector and tensor helicity components and their parametrization depends on the gauge choice.
We will mostly employ the conformal Newtonian gauge where,
\be\label{eq:conformal-gauge}
ds^2 = a^2(\tau) [1+2\Phi(\vec x\,, \tau)]d\tau^2 - a^2(\tau)  [\delta_{ij}(1-2\Psi(\vec x\,, \tau))-h_{ij}(\vec x\,, \tau)]dx^i dx^j\,.
\ee
The metric perturbations are then organized in
\begin{itemize}
\item Scalar $\Phi$ and $\Psi$. For later convience we define the two perturbations
\be
\Theta=\Phi+\Psi,\qquad \Sigma=\Phi-\Psi\,,
\ee
which are sourced respectively by $\delta_{\mu\nu}T^{\mu\nu}$ and by the trace $\eta_{\mu\nu}T^{\mu\nu}$. Notice that $\Sigma$ vanishes in absence of anisotropic stress. We will make use of perturbations in Fourier space, such as
\be
\Theta(\vec x, \tau)=\int \frac{d^4q}{(2\pi)^4}\Theta(\vec q\,, q_0)e^{-i q_0 \tau +i \vec{q}\cdot \vec x}\,.
\ee
The reality of the perturbations implies that the Fourier modes $\Theta(\vec q\,, q_0)^*=\Theta(-q_0,-\vec q\,)$, and similarly for $\Sigma$.

\item Tensor $h_{ij}$. These are transverse traceless perturbation. Here and in the following we use the Fourier and helicity decomposition
\be\label{eq:hij}
h_{ij}(\vec x, \tau)= \sum_{\lambda=\pm}\int \frac{d^4 q}{(2\pi)^4} \varepsilon^\lambda_{ij}(\vec q\,) h_\lambda(\vec q,q_0) e^{-i q_0 \tau +i \vec{q}\cdot \vec x}\,.
\ee
Notice that they are not necessarily on-shell. Reality of the perturbations implies that $h_\lambda^*(q,q_0)=h_\lambda(-q,-q_0)$ provided that $\varepsilon_{ij}^{*\pm}(\vec q\,)=\varepsilon^\mp_{ij}(-\vec q\,)$, and $\varepsilon_{ij}(\vec q)=\varepsilon_{ij}(-\vec q)$. The $\epsilon_{ij}^\pm$ are the two symmetric transverse traceless polarization matrices for a GW of wavenumber $\vec q$, see appendix \cite{Redi:2021ipn}.

\item Vector $W_i$.
We choose a gauge where this are described by the spatial part of the metric as,
\begin{equation}
h_{ij}= \partial_i V_j^T + \partial_j V_i^T \,,\quad\quad \partial^i V_i^T=0\,,
\label{eq:VB}
\end{equation}
that can expanded as,
\be
V_i^T(\vec x\,, \tau) = \sum_{\pm} \int \frac{d^4q}{(2\pi)^4}\epsilon_{i}^{\pm}(\vec q) W_{\pm}(\vec q\,,q_0) e^{-i q_0 \tau +i \vec{q}\cdot \vec x}~~~~~~~~q^i \epsilon_i^{\pm}=0\,.
\ee
\end{itemize}

In this work we will be interested in stochastic backgrounds that are translationally and rotationally invariant. 
The statistical properties of a stochastic variable $X$ are then encoded in their two-point functions at unequal times. 
Invariance under space translations and rotations implies that the two-point function of each helicity perturbation has the form,
\be
\langle X (\vec q,\tau)X(\vec q\,',\tau')\rangle=(2\pi)^3 \delta^3(\vec q + \vec q')\frac{2\pi^2}{q^3}\Delta_{X}(q, \tau,\tau')\,.
\ee
The central object that is needed to compute particle production is the (anti-)Fourier transform of this un-equal time
two-point function. We thus introduce,
\be
\label{eq:DELTA}
\begin{split}
\langle X (\vec q\,, q_0)X(\vec q\,',q_0')\rangle& = (2\pi)^3 \delta^{(3)}(\vec q +\vec q\,')\tilde{\Delta}_{X}(q, q_0, q_0'\,)\,,\\
\tilde{\Delta}_{X}( q, q_0, q_0'\,)&\equiv\int d\tau\int d\tau'\, e^{-i q_0\tau -i q_0'\tau'}\Delta_{X}(q, \tau,\tau')\,.
\end{split}
\ee
In what follows  with abuse of notation we denote the power spectrum in frequency space with the same $\Delta_{X}$ of the power spectrum at unequal time. 
The function $\Delta_X(q,q_0,q_0')$ encodes all information of the two-point function, but only a smaller set of kinematic configurations is relevant for particle production.  
Just by comparing with the master formula in eq.~\eqref{eq:Nparticles}, we see that the relevant quantity is the equal frequency two-point function,
\be\label{eq:building-block}
\langle X (\vec q\,, q_0)X (-\vec q\,, -q_0)\rangle = (2\pi)^3 \delta^{(3)}(0)\frac{2\pi^2}{q^3} \Delta_X (q\,,q_0\,,-q_0)\,,
\ee
Here $(2\pi)^3\delta(0)$ should be identified as the three-dimensional volume factor $V_3$ so that, as expected to an isotropic source, what makes sense is the number density
of produced particles. 

In complete generality from \eqref{eq:Nparticles} one obtains,
\begin{equation}
\frac {d(n_J a^3)}{d q_0}= \frac {c_J}{320\pi^2}  \int_0^{q_0} d(\log q)\Delta_X(q, q_0, -q_0) f_X(q\,,q_0) \,,
\label{eq:densities}
\end{equation}
where the extrema of integration were enforced by $\theta(q^2)$ and we integrated only over positive $q_0$ including a factor 2.
Here the kernel function $f_X(q\,,q_0)$ is a function that depends on the  better spin of the fluctuation that follows from \eqref{eq:Nparticles} when specialized to the corresponding 
(unpolarized) perturbation,
\begin{equation}
f_\Theta(q,q_0)= \frac {q^4}{3}\,,\quad\quad f_h(q,q_0)= \frac {(q_0^2-q^2)^2}{2}\,,\quad\quad f_W(q,q_0)=\frac{q_0^2 q^2 (q_0^2-q^2) }4\,.
\label{eq:kernels}
\end{equation}
We emphasise that any conformally coupled matter or sector, described by a central charge $c_J$, is produced gravitationally with this mechanism.
These formulas only differ for the kernel $f_X(q_0\,,q)$ that multiplies the power spectrum of the relevant perturbation. 
The power spectrum $\Delta_X(q\,,q_0\,,-q_0)$ encodes the spatial and time evolution of the perturbations that we assume to vanish in the far past and in the far future. 
There is then necessarily a period where the fluctuations grows and then eventually decreases. Both these transient epochs can contribute to particle production as we will show in the following.
 
The expression of the kernels  in eq.~\eqref{eq:kernels} make clear that GW in flat space does not produce particles because $q_0^2=q^2$  so that $f_h$ in \eqref{eq:kernels} vanishes. Physically this can be understood as follows. Particle production from inhomogeneities  can be interpreted as decay of perturbations in flat space.  Since GWs have $q_\mu q^\mu=0$ it will not produce particles. 
What  will generate particles is the non-adiabatic process of building up the gravity wave signal and its cosmological evolution. 

The total number of particles is obtained integrating the differential density (\ref{eq:densities}). 
These integrals are finite, for reasonably well behaved fluctuations. Let us discuss the convergence of the integral. 
For tensor perturbations because the kernel grows as $q_0^4$ finiteness of the integral requires that $\Delta_X(q\,,q_0,-q_0)$ goes to zero faster than $1/q_0^5$ for large $q_0$. 
On the other hand for scalar perturbations the kernel is constant in $q_0$ so it is sufficient that the power spectrum goes to zero faster than $1/q_0$.
These conditions are actually automatic if the background is smooth. After all, as per eq.~\eqref{eq:NparticlesWeyl}, particle production is just an integral of the Weyl tensor squared
so that UV divergences cannot arise if the geometry is regular, i.e. if the metric is differentiable.
Indeed for a smooth function $g(x)$ the Fourier transform  goes to zero exponentially at large frequencies.
More in general the asymptotic behaviour is just related to the regularity of the function and of its derivatives
\footnote{Consider a function with continuous n-th derivatives at the origin, 
\begin{equation}
f^{(n)}(t)=\int  \frac{d\omega}{2\pi} (i \omega)^n g(\omega) e^{i \omega t}
\end{equation}
Convergence of the integral requires $g(\omega) \sim 1/\omega^n$.
Therefore the Fourier transform of function $C^{(n)}$ (a function with continuous n-th derivative) must go to zero faster than $1/\omega^{n+1}$ at infinity.}.
In physical systems we expect the geometry away from singularities to be $C^\infty$ so that all the integral will be just finite.
We further elaborate on this point in the appendix with explicit examples. 
We thus emphasize that the particle production leads to a finite result with no need of regularization. 
Moreover the formulas above may lead to the impression gravity wave perturbations are more sensitive to the UV behaviour than scalar perturbations. 
This intuition however turns out to be misleading once we consider how the background is generated. In the next section 
we will see in particular how scalars and tensor perturbation behave similarly in terms of the energy momentum tensor that creates the background.

\subsection{Unequal time correlators: estimates}
\label{sec:estimates}
Particle production from metric perturbations strongly relies on the presence of a non-trivial unequal time correlations $\Delta_X(q,\tau,\tau')$. Having clarified the origin and validity of eq.~\eqref{eq:Nparticles} we can now discuss the impact of realistic perturbations. Let us notice that in \cite{Garani:2024isu,Garani:2025qnm} the main focus was on perturbations with spatial and time dependence generated by the inflaton and then evolving in standard cosmology. Most of the results were therefore based on the time-evolution of perturbations controlled by the Hubble expasions, therefore with a typical momentum scale $\sim H(t)$. Unsurprisingly, particle production was then localized in time and space to scales comparable to the momentum mode of a perturbation experiencing a large time evolution (for example scalar fluctuations re-entering the horizon in radiation domination \cite{Garani:2024isu}). 

We emphasize however, that since a perturbation has to be generated by some microphysics, each cosmological fluctuations in reality will have at least two unavoidable time dependencies: one associated with its build-up, on a time scale $\beta^{-1}$, and the other associated to its damping, often on scales of order $H^{-1}$. It is interesting to explore scenarios where $\beta\gg H$, in search for larger production rates. 

With this mindset, we can now discuss on general grounds what are the expected size of the power spectra $\Delta_X(q,\tau,\tau')$ of eq.~\eqref{eq:DELTA}. Time dependence is essential, otherwise the inclusive formula will contain $\delta(q_0)$ leading to no production on the time-like support. If $\Delta_X(q,\tau,\tau')$ enjoys time-translational invariance, in addition to the spatial one, it follows that the production rate will be extensive in time (a situation that we will encounter in section \ref{sec:GFI}). However in many other context we do not expect time-translation be preserved, but rather the perturbations will be coherent in time, so that the power spectrum and its fourier transform read
\begin{equation}\label{eq:coherent}
\Delta(q,\tau,\tau')=A(q) T(q,\tau)T^*(q,\tau')\,,\quad \Delta(q,q_0,-q_0)=A(q) |\mathcal{I}(q\,,q_0)|^2\,,
\end{equation}
where $\mathcal{I}(q\,,q_0)\equiv \int d\tau \exp(-i q_0 \tau) T(q,\tau)$. For coherent sources, given the expressions in \eqref{eq:kernels}, the particle production abundances can be constructed with the moments computed for time-like configurations
\begin{equation}
g_n(q)\equiv 2\int_q^\infty q_0^{2n} |{\cal I}(q,q_0)|^2 dq_0,
\label{eq:moments}
\end{equation}
leading to 
\be
n_Ja^3= \frac{c_J}{1280\pi^2} 
\begin{cases}
2 \int  d(\log q)  A(q) \left[g_2(q)-2 q^2 g_1(q)+q^4 g_0(q)\right],&{\rm tensor,}\\
2 \int  d(\log q)  A(q) \left[q^4 g_0(q)\right], & {\rm scalar,} \\
2 q^2 \int  d(\log q)  A(q) \left[g_2(q)- q^2 g_1(q)\right] & {\rm vector}.
\end{cases}
\ee
These moments contain energy scales associated to both $\beta$, the build-up, and $H$, the cosmological redshift. They are generated by the non-adiabatic time evolution of the background, a key ingredient for particle production. As we are going to stress in this work, the non-adiabaticity may come from cosmological expansion as it may come from physics inside the horizon. The two have qualitatively different behavior that we are going to explain.

\paragraph{Production inside the horizon}~\\
 For the production inside the horizon we can in first approximation neglect the cosmological evolution at production and work in flat space\footnote{The momenta in the estimates based on flat space throughout the paper are physical rather than co-moving.}. 
 The main feature we are interested in is the time scale over which the gravity wave signal is built. Following the standard notation of first order phase transitions we denote this
as $\beta^{-1}$. As explicit example we can choose to approximate the transfer function on the time scale of its non-adiabatic evolution as
\begin{equation}
T(q,\tau)=\frac 1 2 [1+\tanh(\beta(\tau)) ]\longrightarrow {\cal I}(q_0\,,q)=-\frac{i }{2\beta \sinh \frac{q_0}{2 \beta }}\,,
\label{eq:tanh}
\end{equation}
where we neglected $\delta(q_0)$ in the Fourier transform irrelevant for particle production. 
This corresponds to a perturbation built over a time $\beta^{-1}$.
Since the transfer function is $C^\infty$ its Fourier transform is exponentially suppressed for $q_0\to \infty$.
This guarantees that the integral is finite. The integral can be carried out analytically.
In the limit of a fast transition, i.e. $\beta \gg q$ one finds,
\begin{equation}
g_2 = \frac 4 {15}\pi \beta^3 +\dots\,,~~~~~~~~~~g_1=-\frac 1 3 \pi \beta+\dots \,,~~~~~~~~~ g_0=\frac 1 q+\dots
\end{equation}
Different transfer functions would give different coefficient but same parametric dependence on $\beta$
that follows essentially from dimensional analysis. Notice that each perturbation will evolve differently after the build-up when the source is no longer active, following their typical  mostly adiabatic free evolution as in standard cosmology (see eq.s~\eqref{eq:MB}). Particle densities are then typically created at a scale factor $a_*$, and then redshift with the volume exploiting particle number conservation. To leading order today's abundance is found to be
\[
n_J \sim  \frac{c_J a_*^3}{1280\pi^2} 
\begin{cases}
  \beta^3 \int d \log q  A_2(q),&{\rm tensor,}\\
   \int d \log q  q^3 A_1(q), & {\rm scalar,} \\
 \beta^2  \int d \log q  q^3 A_0(q),& {\rm vector}.
\end{cases}
\]
These formulas might give the impression that GW lead to larger abundance of particles for fast transitions.
In realistic situations however the peak of the power spectrum also occurs on spatial momenta of order $q_*\sim \beta$. 
As a consequence the precise abundance depends on the details on the transfer function, but production from scalar and tensor perturbations inside the horizon gives overall similar estimates of the type
\begin{equation}
n_J \sim 10^{-3} c_J \left(\beta a_*\right)^3 A(q_*) = 10^{-3} c_J \left( \frac{\beta}{H_*}\right)^3 a_*^3 H_*^3 A(q_*)
\label{eq:estimate}
\end{equation}
From this we see that production inside the horizon from a perturbation of the type \eqref{eq:tanh} depends on the combination scale factor times Hubble at the epoch of production $a_* H_*$. The other quantities do not depend on the cosmological evolution since both $\beta/H_*$ and $A(q_*)$ are related to intrinsic parameters of the source leading to perturbations.
Notice that the quantity $a^3 H(a)^3$ is maximal at the end of inflation, therefore the largest production happens right after the end of inflation when the comoving Hubble radius starts increasing, for a fixed $A(q_*)$.

To compare with other production mechanisms we can define the yield $Y\equiv n_J/s_0$, where $s_0$ is the today's entropy density. If $a_*$ is an epoch during radiation dominance, we estimate
\be\label{eq:yield}
Y_J\big|_{\rm rad.} \sim 10^{-2} c_J \left(\frac{\beta}{H_*}\right)^3 \left(\frac{T_*}{M_{\rm Pl}}\right)^3 A(q_*)\,.
\ee
If we allow for a finite duration of reheating -- matter dominated -- leaving the possibility that $a_*<a_R$, where $a_R$ denotes the scale factor at the onset of radiation dominance with temperature $T_R$. 
The yield becomes
\be\label{eq:yieldRH}
Y_J\big|_{\rm reheating} \sim 10^{-2} c_J \left(\frac{\beta}{H_*}\right)^3 \left(\frac{T_R}{M_{\rm Pl}}\right)^3  \left(\frac {a_R}{a_*}\right)^{3/2}A(q_*)\,.
\ee
For fixed $\beta/H_*$ and $A(q_*)$, the abundance $Y_J$ is larger if it is produced during reheating despite the fact that entropy injection dilutes the number density. This happens because $a_*^3 H_*^3$ compensate for the dilution.  Thanks to this, particle production is particularly efficient when it takes place right after the end of inflation. The boost factor in that case is $(a_*/a_R)^{3/2}$, related to the evolution of Hubble as $a^{-3/2}$ in a phase of matter dominance.

The expert reader may have noticed that the formulas above look similar to gravitational freeze-in production of DM from the Standard Model thermal bath. 
For freeze-in, which is a thermal process, the abundance produced during reheating is negligible and thus it is mostly determined 
by the reheating temperature. In that case one finds  \cite{Redi:2020ffc}
\begin{equation}
Y_{\rm GFI}\approx 5\times 10^{-6} c_J \left(\frac{T_R}{M_{\rm Pl}}\right)^3 \,.
\end{equation}
As we will show in the next section \ref{sec:GFI} the similarity with eq.~\eqref{eq:yield} is not accidental and indeed we will be able to derive gravitational freeze-in
from particle production due to inhomogeneities.

\paragraph{Production from cosmological evolution}~\\
Cosmological evolution also leads to particle production. This is the main mechanism that was studied in \cite{Garani:2025qnm,Garani:2024isu}
where perturbations produced during inflation were considered to populate DM. 
Usually perturbations start evolving when the modes re-enter the horizon and the evolution is controlled by the Hubble parameter. 
Note that also GW can produce particles in this case because their wave-function is not purely a plane wave. 
This situation can again be captured with the simple transfer function (\ref{eq:tanh}) with the only difference that  $\beta\sim -H$ because the perturbation dies out rather than growing. 
This however makes no difference for what concerns particle production so that in particular the estimate (\ref{eq:estimate}) applies.
Here however differences will appear between tensor and scalar perturbations.
Tensor perturbations start to evolve as soon as they are produced so that production is maximal at the initial time where $H a$ is maximal.
Given that  $\beta \gg H$ for production inside the horizon it follows that the cosmological evolution will typically give a small correction to the number of particles produced. 

However, there is an interesting exception for scalar perturbations that re-enter the horizon in a matter dominated era, as in that case  
the perturbations remain constant even inside the horizon. This situation is realized during reheating where the energy density approximately redshifts as matter.
In that case the evolution of scalar perturbations takes place at the onset of radiation domination so that the particles are produced well after perturbations have re-entered
the horizon.  The delayed production generates a larger abundance, unaffected by dilution approximately given by (\ref{eq:yieldRH}) with $a_*=a_R$, i.e. the scale factor at reheating time.

The precise evolution of scalar perturbations can be computed following \cite{Garani:2025qnm}, see appendix for a more general derivation.
In absence of sources and neglecting anisotropic stress, scalar perturbations satisfy the simple  equation,
\begin{equation}
\Theta''(q\,,\tau)+3 \frac {a'}{a}(1+c_s^2) \Theta'(q\,,\tau) + c_s^2 q^2  \Theta(q\,,\tau)=0\,,
\label{eq:Phi}
\end{equation}
where $c_s$ is the adiabatic speed of sound.
During radiation $c_s^2=1/3$ this the equation of a damped harmonic oscillator, leading to $\Theta \propto 1/(c_s q\tau)^2$ so that scalar perturbations become quickly suppressed inside the horizon. 
On the other hand during matter domination the solution is,
\begin{equation}
\Theta_{\rm mat}= c_0+ \frac {c_1}{\tau^5} \,.
\end{equation}
As a consequence scalar perturbations do not decay and remain constant even for modes that are already deep inside the horizon.
Denoting with $A_0(q)$ the amplitude of the equal time power spectrum produced during the reheating phase one obtains the abundance
\begin{equation}
na^3 \approx 5 \times 10^{-4}\times c_J   \int d (\log q) q^3 A_0(q) \,.
\label{eq:abinflation}
\ee
The result is only approximate because the integral mildly depends on $\tau_R$.
This should be compared with the result from the production of the perturbation during the build-up phase in eq.~\eqref{eq:estimate}.
The two contributions can be comparable. 


\section{Particle Production inside the horizon}\label{sec:micro}
The formulas of section \ref{sec:review} acquire a particular simple if the pertubation of the metric is generated inside the Hubble horizon so that the cosmological 
evolution can in first approximation be neglected at production. In this section we will relate the particle production to the two-point function of the energy momentum tensor
that produce the perturbations.

\subsection{Production of matter and GW}
At distances shorter than the Hubble radius one can treat the system in  flat space and make a stronger connection with the microphysics responsible 
for the generation of the fluctuations.  This will also uncover the correlation between particle and GW production. 

To be concrete, we consider a sector $\Scal$ with non-trivial dynamics (be it a phase transition, a thermal sector, etc...) happening on scales shorter than the Hubble radius. 
Its dynamics provides a stress-energy tensor $T_{\mu\nu}^{\Scal}(\vec x, \tau)$, that is responsible for the generation of the perturbations of eq.~\eqref{eq:metric}, that we label as $h_{\mu\nu}^{\Scal}$, to make their origin manifest. Inside the horizon, $h_{\mu\nu}^{\Scal}$ is just the solution of linearized Einstein's equations in flat space,
 therefore the computation relies on determining the background geometry $\bar g_{\mu\nu}=\eta_{\mu\nu}+h_{\mu\nu}^{\Scal}$ at leading order in the source. 
In the harmonic gauge, we can write
\be\label{eq:h-from-T}
\Box h_{\mu\nu}^{\Scal}=\frac{2}{\Mpl^2}T_{\mu\nu}^{\Scal}\,,\quad\quad h_{\mu\nu}^{\Scal}(q)=\frac{-2}{(q^2+i\epsilon q_0) \Mpl^2}T_{\mu\nu}^{\Scal}(q)\,,
\ee
where we  the $i \epsilon$ prescription selects the retarded Green's function. 
Plugging the solution above into eq. (\ref{eq:Nparticles}) we thus obtain a formula for the particles produced in 
terms of the energy momentum tensor of the source.

To see how this works explicitly let us now generalize the derivation of the effective action to include also fluctuations of the metric on top of the classical background. 
Thus we consider as quantum dynamical fields both a generic CFT and gravitons $\gamma_{\mu\nu}$. Their action in the background of \eqref{eq:h-from-T} is
\be
S[\text{CFT};\gamma_{\mu\nu}; \text{$\mathcal{S}$}]= S_{\text{CFT}}|_\eta + \int \frac{1}{2} h_{\mu\nu}^{\Scal}T^{\mu\nu}_{\rm CFT}+ \int \frac{1}{2} \frac{\gamma_{\mu\nu}}{\Mpl} T^{\mu\nu}_{\Scal} + (\text{free $\gamma_{\mu\nu}$ action}) +\cdots
\ee
where to leading order all indexes are lowered and raised with $\eta_{\mu\nu}$.

As in section \ref{sec:review} the 1PI effective action is obtained in terms of $T$-ordered two-point functions $ \langle T^{\mu\nu}_{\rm CFT}(x)T^{\rho\sigma}_{\rm CFT}(y) \rangle$ and  $\langle \gamma^{\mu\nu}(x)\gamma^{\rho\sigma}(y) \rangle$. The former is again fixed by conformal symmetry as in \eqref{eq:2pointTT}, while the latter is the graviton propagator. The 1PI effective action -- in the background of $\Scal$ -- is then 
\be\label{eq:1PIS}
\Gamma_{\rm 1PI}[\Scal]=\frac{i}{8\Mpl^2}\int \frac{d^4 q}{(2\pi)^4} T_{\mu\nu}^{\Scal}(q) T_{\rho\sigma}^{\Scal}(-q)\left[\frac{4}{q^4 \Mpl^2} \langle T^{\mu\nu}_{\rm CFT}(q)T^{\rho\sigma}_{\rm CFT}(-q)\rangle + \langle \gamma^{\mu\nu}(q) \gamma^{\rho\sigma}(-q)\rangle \right]\,.
\ee
where,
\begin{equation}
\langle \gamma^{\mu\nu}(q) \gamma^{\rho\sigma}(-q)\rangle= \frac {4i}{q^2+i \epsilon} P_{\mu\nu\rho\sigma}\,,~~~~~~~~P^{\mu\nu\rho\sigma}\equiv\frac12(\eta^{\mu\rho}\eta^{\nu\sigma}+\eta^{\mu\sigma}\eta^{\nu\rho}-\eta^{\mu\nu}\eta^{\rho\sigma})
\end{equation}
and the two-point function of $T^{\mu\nu}_{\rm CFT}$ is given in (\ref{eq:2pointTT}). The factor of 4 in the graviton propagator arises from the fact that the coupling to $T_{\mu\nu}$ corresponds to a non-canonically normalized $\gamma_{\mu\nu}=2 \gamma_{\mu\nu}^{\rm can.}$. Crucially here the graviton propagator here has the $i\epsilon$ Feynman prescription for T-ordering. 
The different tensor structure is due to the fact that the metric perturbations are not conformally coupled.
As expected, both the new sector (CFT) and gravitons contribute to the effective action.

By taking the imaginary part of $\Gamma_{\rm 1PI}[\Scal]$ in eq.~\eqref{eq:1PIS}, we compute the inclusive production probability of the CFT sector and gravitons. 
Using the identity $1/(q^2+i \epsilon)={\rm PV}(1/q^2)-i \pi \delta (q^2)$ the inclusive probability is then
\be\label{eq:imaginary1PIS}
2\mathrm{Im}[\Gamma_{\rm 1PI}[\Scal]]=-\frac{1}{\Mpl^2}\int \frac{d^4q}{(2\pi)^4}\langle T_{\mu\nu}^{\Scal}(q) T_{\rho\sigma}^{\Scal}(-q)\rangle_{\rm in.} \left[ \frac{c_J}{7680\,\pi\,\Mpl^2} \frac 1 {q^4}\Pi^{\mu\nu\rho\sigma}\theta(q^2) + P^{\mu\nu\rho\sigma}\pi \delta(q^2) \right]
\ee
This formula is completely general. The first term describes particle production while the second corresponds to production of gravitons\footnote{
Upon using the on-shell condition $q^2=0$, and the conservation of $T_{\mu\nu}$, one can show that $\langle T_{\mu\nu}^{\Scal}(\vec q\,,q) T_{\rho\sigma}^{\Scal}(-\vec q\,,-q)\rangle P^{\mu\nu\rho\sigma}=\langle T_{ij,\rm TT}^{\Scal}(\vec q\,,q) T_{ij,\rm TT}^{\Scal}(-\vec q\,,-q)\rangle$, where TT now refers to the transverse traceless spatial components of the metric. This allows to compute 
the inclusive production of GW without explicitly imposing the TT projection on the source.}. GWs are produced provided that the source has a non-zero anisotropic stress with time dependence on a null-like support.  Particle production instead requires less: only time dependence of the source on a time-like support.  

The two-point function of the initial source (which need not to be time-orded) is usually computed as an expectation value on the initial state, therefore we have included the notation $\langle\cdots \rangle_{\rm in.}$. For our derivation, the initial state can be generic (quantum, statistical, thermal, etc.) provided that we are doing a sub-horizon calculation.

The expression above invite for a derivation in terms of Feynman diagrams as shown in Fig. \ref{fig:feynman}.
Particle production corresponds to the inclusive production of the spectator field through graviton exchange.
On the other hand production of GW is associated to the single production.  Both quantities can be obtained 
as the imaginary part of the associated forward diagram.

\paragraph{Differential number and energy densities}~\\
From the previous formulas we can also derive the expression for the differential distributions.
Assuming an isotropic source we obtain, 
\begin{equation}\label{eq:inclusive-formulas-FLAT}
\begin{split}
\frac{dN}{dq_0 d \log q}&=\frac 1 {\Mpl^4} \frac{c_J}{640\pi^2 } \frac{q^3}{2\pi^2}\times\left[ \langle T^{\Scal}_{\mu\nu}(q)T^{\Scal\,\mu\nu}(-q)\rangle-\frac13 \langle T^{\Scal}(q)T^{\Scal}(-q)\rangle \right]\times  \theta(q_0)\theta(q_0^2-q^2)\,,\\
\frac{dE_{\rm gw}}{d\log q}&=\frac{1}{\Mpl^2}\frac{q^3}{2\pi^2} \times \langle T_{ij,\rm TT}^{\Scal}(q) T_{ij,\rm TT}^{\Scal}(-q)\rangle\Big|_{q_0=|q|} \,.
\end{split}
\end{equation}
where the first equation refers to the number of events while in the second we have integrated $\delta(q^2)$ for GW and reported the energy density. 

Importantly both CFT and GW production are controlled by the two-point function of the stress-energy tensor of the source for the kinematic configuration $q_0>q$ and $q_0=q$ respectively. In presence of extensive sources -- as when there is spatial translational invariance -- one can definite the number and energy density.

The formula for gravitational wave production agrees with the standard formulas in literature derived classically, see \cite{Caprini:2018mtu}. 
Here we have shown that it can be derived using the quantum formalism of the effective action. 

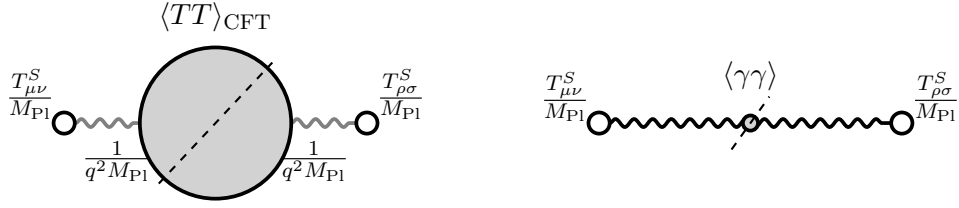
\begin{figure}[t]
\centering

\begin{tikzpicture}[line width=1.4pt, scale=1]
  \path[use as bounding box] (-2.7,-1.15) rectangle (2.7,1.15);
 \draw[decorate, color=gray, decoration={snake, amplitude=1.5pt, segment length=7pt}]
    (-1.85,0) -- (-1,0);
  \draw[decorate, color=gray, decoration={snake, amplitude=1.5pt, segment length=7pt}]
    (1,0) -- (1.85,0);

  \draw (-2,0) circle (0.14);
  \draw (2,0) circle (0.14);

  \node at (-2.45,0.38) {$\frac{T_{\mu\nu}^{S}}{\Mpl}$};
  \node at (2.45,0.38)  {$\frac{T_{\rho\sigma}^{S}}{\Mpl}$};

  \draw[fill=gray!35, draw=black] (0,0) circle (1cm);
  \node at (0,1.4) {$\langle T T\rangle_{\rm CFT}$};

  \node at (-1.3,-0.5) {$\frac{1}{q^2\Mpl}$};
  \node at (1.3,-0.5)  {$\frac{1}{q^2\Mpl}$};

  \draw[dashed, thick] (-0.75,-0.8) -- (0.75,0.8);
\end{tikzpicture}
\qquad\qquad
\begin{tikzpicture}[line width=1.4pt, scale=1]
  \path[use as bounding box] (-2.7,-1.15) rectangle (2.7,1.15);

  \draw (-2,0) circle (0.14);
  \draw (2,0) circle (0.14);

  \node at (-2.45,0.35) {$\frac{T_{\mu\nu}^{S}}{\Mpl}$};
  \node at (2.45,0.35) {$\frac{T_{\rho\sigma}^{S}}{\Mpl}$};

  \draw[decorate, decoration={snake, amplitude=1.5pt, segment length=7pt}]
    (-1.86,0) -- (1.86,0);

  \node at (0,0.62) {$\langle \gamma \gamma \rangle$};

  \draw[fill=gray!35, draw=black] (0,0) circle (.1cm);

  \draw[dashed, thick] (-0.25,-0.35) -- (0.25,0.35);
\end{tikzpicture}

\caption{\label{fig:CFT-gravity-sources} Contributions to the $\Gamma_{\rm 1PI}$ effective action sourced by the sector $\Scal$. On the left, leading order contribution from the two-point functions of the CFT in the metric background sourced by $T_{\mu\nu}^{\Scal}$. On the right, leading order tree-level contribution from the graviton propagator in the background of the source $T_{\mu\nu}^{\Scal}$.}
\label{fig:feynman}
\end{figure}

\paragraph{Computation of the two-point function on the initial state }~\\
Particle and GW production depend on the expectation value of $\langle T^{\Scal}_{\mu\nu}(q) T^{\Scal}_{\rho\sigma}(-q)\rangle_{\rm in.}$  properly contracted with the corresponding projectors. Such a function can be computed classically, thermally or quantistically provided we know the density matrix of the initial state of the source. Conceptually this is no different from computing expectation value tracing over a density matrix $\hat\rho$, namely $\langle \cdot \rangle\equiv \mathrm{tr}[\hat\rho \cdots]$, for microphysics happening inside the horizon. For example, for computation in the Bunch-Davies vacuum we have $\rho=|\text{0}\rangle \langle \text{0}|$, for a thermal state $\hat\rho=e^{-\beta \hat H}$, or for a 2-particle initial state $\hat\rho=|p_1,p_2; q=p_1+p_2\rangle \langle p_1,p_2; q=p_1+p_2|$. Or in general, the two-point function correspond to a classical and conserved stress-energy tensor. We consider explicit applications in section \ref{sec:GFI}, gravitational freeze-in of DM from the thermal SM bath mediated by graviton exchange, and \ref{sec:PT}, first order phase transitions with generation of anisotropic stress.

\subsection{Production from Scalar, Vector and Tensor perturbations}

From the derivation so far we have seen that the production of particles and GWs takes a universal form in terms of the two-point function of the stress-energy tensor of the source. 
In particular eq.~\eqref{eq:inclusive-formulas-FLAT} is really everything we need to compute inclusive production rates.

However, to make contact with the nomenclature often used in the literature, in this subsection we are going to derive more exclusive formulas that will relate the stress-energy two-point function of the source with the power spectrum of the generic perturbations as introduced in section \ref{sec:review}, see eq.s~\eqref{eq:DELTA}, for the typical choice of the conformal-newtonian gauge. For example we will see that the source for GWs corresponds to the usual TT-gauge tensor contribution. 

In particular having outlined the derivation inside the horizon, we can compute the $\Delta_X(q,q_0,-q_0)$ for $X=\Theta, h, W$ including the universe expansion and connecting them to the two-point function of the source. By doing so, we loose the compactness of eq.~\eqref{eq:inclusive-formulas-FLAT} but we could use the general formulas of section \ref{sec:review} informed by microphysical processes. For doing so, we use the linearized Einstein's equations for the conformal-newtonian gauge following the standard derivation of Ma and Bertschinger \cite{Ma:1995ey}. In general, the components of the stress-energy tensor of the source, $T^{\mu\nu}$, relevant for us, are encoded in
\be\label{eq:components}
T^{0}_{\,\,0}\equiv \rho, \quad T^{i}_{\,\, i}\equiv -3p,\quad T^0_{\,\, i}, \quad \Sigma_{ij}\equiv T_{ij}-\frac13 \delta_{ij}T^k_k  \,,
\ee
where we have defined the anisotropic stress-tensor $\Sigma_{ij}$ (traceless). Expanding around an FLRW background, in the gauge (\ref{eq:conformal-gauge}), the linearised Einstein's equations, in Fourier space $e^{i \vec x \cdot \vec q}$, take the form
\begin{subequations}\label{eq:MB}
\begin{align} 
-q^2\Psi -3\mathcal H\left(\Psi'+\mathcal H\Phi\right) &= \frac{a^2}{2M_{\rm Pl}^2}\,\delta\rho ,
\label{eq:drho}\\
\Psi'' +\mathcal H\left(\Phi'+2\Psi'\right) +\left(2\mathcal H'+\mathcal H^2\right)\Phi -\frac{q^2}{3}\left(\Phi-\Psi\right)
&= \frac{a^2}{2M_{\rm Pl}^2}\,\delta p ,
\label{eq:dp}
\\
q^2\left(\Phi-\Psi\right)&= \frac{3a^2}{2M_{\rm Pl}^2}\, \left(\bar\rho+\bar p\right)\sigma ,
\label{eq:sigma}
\\
h_{ij}''+ 2 \mathcal{H}h_{ij}' + q^2 h_{ij}
&=
\frac{2}{M_{\rm Pl}^2}\,T_{ij}^{\rm TT} ,
\label{eq:tensor}
\\
-q^2 W_i^{T\,\prime}
&=
\frac{2}{M_{\rm Pl}^2}\,
a^2\left(\bar\rho+\bar p\right)\delta T_i^{0\,{\rm T}} .
\label{eq:vector}
\end{align}
\end{subequations}
We have introduced the conformal Hubble scale $\mathcal{H}\equiv a'/a^2$, while for tensor perturbations we conveniently introduced the rescaled perturbation $H_{ij}= a h_{ij}$. In this set of equation, the average energy $\bar \rho$ and pressure $\bar p$ appear as well as the scalar anisotropic stress $\sigma$. Explicitly, the matter sources appearing on the righ-hand side of the above equations are extracted from eq.~\eqref{eq:components} by means of the spatial projector $P^{\rm T}_{ij}(\vec q\,)\equiv \delta_{ij}-\frac{q_i q_j}{q^2}$.
This allows us to identify the following scalar, vector and tensor sources
\begin{equation}
   T_{ij}^{\rm TT}=(P^{\rm T}_{ik}P^{\rm T}_{jl}
-\frac12 P^{\rm T}_{ij}P^{\rm T}_{kl})\Sigma_{kl}\,,\quad T^{0,\rm T}_{\,\,i}= P_{ij}^{\rm T}T^{0}_{\,j}\,,\quad 
 (\bar\rho+\bar p)\sigma= P_{ij}^{\rm T} \Sigma_{ij} \,.
 \end{equation}

\subsubsection{Flat space limit}
While the formulas above can be solved in general they become extremely simple in the sub-horizon regime where we neglect the 
expansion of the universe, $a\to 1$ and $\mathcal{H}\to 0$, and they reduce to \eqref{eq:h-from-T} when we switch it from deDonder to conformal-newtonian gauge. The Fourier space solutions for $\Theta, V, h$ are
\begin{eqnarray}
\label{eq:TTheta}
\Theta(\vec q\,, q_0)&=& \frac{-1}{q^2\Mpl^2} T_{\Theta}(\vec q\,, q_0)\,,\quad\quad T_{\Theta}\equiv \delta T^0_{\,\, 0} -\frac 3 2  (\rho+p)\sigma\,,\\
W_j(\vec q\,, q_0)&=& \frac{-2i}{q_0 q^2 M_{\rm Pl}^2}  T^{0\,\rm T}_j(\vec q\,, q_0)\,,\\
h_{ij}(\vec q\,, q_0)&=&\frac{-2}{(q_0^2-q^2)\Mpl^2}T_{ij}^{\rm TT}(\vec q\,, q_0)\,.
\end{eqnarray}
In the flat space limit there are just three possible source functions.\footnote{For non-conformally coupled massless particles the derivation can be immediately extended to $\Sigma=\Phi-\Psi$.} And by taking the square of the above solutions, we can compute the power spectrum of all the perturbations. Using some standard notation, we define the two-point function of the unpolarized source as
\begin{eqnarray}
\langle T_\Theta(\vec q, q_0) T_{\Theta}(\vec q\, ',q_0')\rangle&=&(2\pi)^3\delta^3(\vec q + \vec q\,') \Pi_{\Theta}(q, q_0, q_0')\,,\\
\langle T^0_i(\vec q, q_0) T^0_i(\vec q\, ',q_0')\rangle&=&(2\pi)^3\delta^3(\vec q + \vec q\,')  \Pi_{W}(q, q_0, q_0')\,,\\
\langle T^{\rm TT}_{ij}(\vec q, q_0) T^{\rm TT}_{ij}(\vec q\, ',q_0')\rangle&=&(2\pi)^3\delta^3(\vec q + \vec q\,') \Pi_{\rm GW}(q, q_0, q_0')\,.
\end{eqnarray}
Notice that for vector and tensor sources the above spectra are related to helicity components (see the decomposition in eq.~\eqref{eq:hij}). We these expression we can then complete the relation between perturbations and sources in flat space as summarized by table \ref{tab:summary}.
From table \ref{tab:summary} we notice that all the particle abundances are similar once expressed in terms of the two-point function of the corresponding source $\Pi(q,q_0,-q_0)$. This explains the origin of the different kernels in (\ref{eq:kernels}). Once expressed in terms of the energy momentum 
tensors there is no structural difference between different perturbations. In particular the abundance remains finite even 
if the energy momentum tensor is discontinuous.

\begin{table}[t]
\centering
\renewcommand{\arraystretch}{2.2}
\begin{tabular}{c|c|c}
\hline
Type 
& Power spectrum 
& Abundance $\displaystyle\frac{dn_J}{dq_0}$
\\
\hline
$\Theta$
&
$\displaystyle 
\Delta_\Theta(q,q_0,q_0')
=
\frac{q^3}{2\pi^2}
\frac{1}{M_{\rm Pl}^4 q^4}
\Pi_\Theta(q,q_0,q_0')$
&
$\displaystyle
\frac{c_J}{960\pi^2} \frac{1}{M_{\rm Pl}^4} \int_0^{q_0} \frac{dq}{q}\, \frac{q^3}{2\pi^2} \Pi_\Theta(q,q_0,-q_0)$
\\
\hline
$V_i$
&
$\displaystyle
\Delta_V(q,q_0,q_0')
=
\frac{q^3}{2\pi^2}
\frac{2}{M_{\rm Pl}^4 q_0^2 q^4}
\Pi_V(q,q_0,q_0')$
&
$\displaystyle
\frac{c_J}{1280\pi^2} \frac{1}{M_{\rm Pl}^4} \int_0^{q_0} \frac{dq}{q}\, \frac{q_0^2-q^2}{q^2} \frac{q^3}{2\pi^2} \Pi_V(q,q_0,-q_0)$
\\
\hline
$h_{ij}$
&
$\displaystyle
\Delta_h(q,q_0,q_0') = \frac{q^3}{2\pi^2} \frac{1}{M_{\rm Pl}^4(q_0^2-q^2)^2} \Pi_{\rm GW}(q,q_0,q_0')$
&
$\displaystyle \frac{c_J}{640\pi^2} \frac{1}{M_{\rm Pl}^4} \int_0^{q_0} \frac{dq}{q}\, \frac{q^3}{2\pi^2} \Pi_{\rm GW}(q,q_0,-q_0)$
\\
\end{tabular}
\caption{Flat-space relations between metric perturbations, their power spectra, and particle production densities. The final abundances are computed with the master formula of eq.~\eqref{eq:densities} and the expressions of the power spectra of the perturbations derived in this section.  For vector and tensor perturbations we consider an unpolorized background and the number densities include a factor 2 multiplicity. }
\label{tab:summary}
\end{table}

\subsection{Correlation between DM abundance and GWs}
The connection unveiled by our derivation relates the spectrum of the GW produced by the phase transition to the DM abundance (or in general any relic abundance). This is due to the fact that they both depend on the power spectrum $\Pi_{\rm GW}(q,q_0,-q_0)$, although sampled in different kinematic regions. 
Explicitly, taking into account cosmological redshift in (\ref{eq:inclusive-formulas-FLAT}), for tensor perturbations we find,
\begin{eqnarray}
\frac{d \rho_{\rm GW} }{d\log q} &=&\left(\frac{a_*}{a_0}\right)^4\frac{1}{\Mpl^2} \frac{q^3}{2\pi^2} \Pi_{\rm GW}(q,q,-q)\,,\\
\frac{d n_{\rm DM}}{dq_0 d\log q} &=& \left(\frac{a_*}{a_0}\right)^3 \frac{c_{\rm DM}}{640\pi^2} \frac{1}{M_{\rm Pl}^4}  \frac{q^3}{2\pi^2}\Pi_{\rm GW}(q,q_0,-q_0)\theta(q_0^2-q^2)\theta(q_0)\,.
\label{eq:abGWPP}
\end{eqnarray}
where in the second formula we assume that each event produces two DM particles.
Assuming the integral over $q_0$ is saturated at $q$ these formulas imply a relation between the abundance of GW
and the one of particles. In particular for a spectrum peaked at $q_0\sim \beta$ we find
\begin{equation}
Y_{\rm DM}\equiv \frac{n_{\rm DM}}s \sim 5\times 10^{-4} c_{\rm DM} \left(\frac {\beta}{H_*}\right) \left( \frac {T_*}{M_{\rm Pl}}\right)^3\frac {\Omega_{\rm GW}}{\Omega_\gamma}
\end{equation}
where $\Omega_\gamma= 5 \times 10^{-5}$ and we assumed the production to take place during radiation domination. 
This estimate can be checked in explicit examples such as phase transitions, see section \ref{sec:PT}.

The energy fraction of DM is given by $\Omega_{\rm DM}=2.7 \cdot 10^8 Y_{\rm DM} M/{\rm GeV}$.
From a given GW background we find that the DM abundance is reproduced for, 
\be
M \sim 10^6\, \mathrm{GeV} 
\left(\frac{10^{-7}}{\Omega_{\rm GW}}\right)
\left(\frac{10^{15}\mathrm{GeV}}{T_*}\right)^3\left(\frac 4 {c_{DM}}\right)\left(\frac{H_*}{\beta}\right)
\,.
\ee
The peak frequency of the GW spectrum today is also determined as $f_0\sim \beta a_*$, see section \ref{sec:PT}.

If the production happens during reheating it will be further enhanced for fixed $\Omega_{\rm GW}$. This happens since $M\propto a_*$ once the DM abundance is fixed. However a phase of matter domination will reduce also $\Omega_{\rm GW}$ from its maximal value of $\approx 10^{-7}$ by an additional redshift of $(a_*/a_R)^4$ where $R$ stands for the onset of radiation dominance where the usual estimates applies. Physically, since this production happens inside the horizon, it will be diluted by any additional redshift originating from a primordial phase of matter dominance while the universe is reheating.

\section{Gravitational Freeze-in}\label{sec:GFI}
An application of our formalism inside the horizon is the calculation of gravitational freeze-in of DM. This process corresponds to the production of DM from the SM thermal bath from $2\to 2$ scatterings mediated by an $s$-channel exchange of gravitons. This type of freeze-in was first introduced in \cite{Garny:2015sjg}, where the Boltzmann equations for DM production were solved with annihilation cross-sections computed exclusively for each channel. The resulting DM mass is highly sensitive to the reheating temperature of the universe, and we do not reiterate its derivation here. In  \cite{Redi:2020ffc,Redi:2021ipn} the same calculation was done by exploiting the approximate scale invariance of the SM and DM sector, relating the inclusive annihilation rate to the central charges of the dark and SM sectors. So far in the literature, gravitational freeze-in was computed by the thermal average of the (inclusive) cross-section.

In this section we emphasize that the same exact result as in \cite{Redi:2020ffc} can be derived by treating the thermal production of gravitationally coupled DM as a specific case of particle production inside the horizon. Our aim here is to compute the rate (per unit time and volume) of the thermally average inclusive production of DM (treated as relativistic ad production, and therefore only described by its central charge). The two possible computations are depicted in figure \ref{fig:GFI}.

Doing the calculation in flat space is a very good approximation since the thermal production is happening on short scales $T^{-1}\ll H^{-1}$. For gravitational freeze-in the temperatures are usually so large that also the source $\Scal$ (the SM at finite temperature) can be approximated as a free relativistic sector (CFT-like). This amounts to neglect the trace of the stress-energy tensor of the source $T^{\Scal}\approx 0$. The use of such a formula implies that we are considering the generation of metric fluctuations $h_{\mu\nu}^{\Scal}$ due to the thermal bath linking them to the stress-tensor of the source. As far as we are aware, this approach was never used in the literature for the calculation of the freeze-in rates.

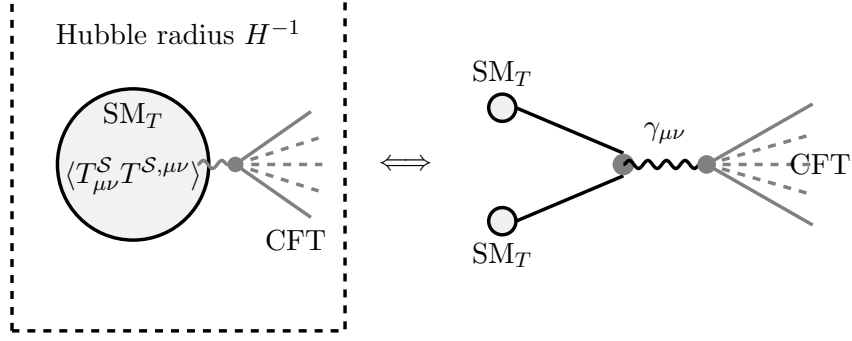
\begin{figure}[t]
\centering
\begin{tikzpicture}[line width=1.3pt, scale=1]
  \path[use as bounding box] (-2.6,-2.2) rectangle (3.6,2.2);
  \draw[dashed] (-2.2,-2.2) rectangle (2.2,2.2);
  \draw[fill=gray!10, draw=black] (-0.6,0) circle (1.0cm);
  \node at (0,1.75) {Hubble radius $H^{-1}$};
  \node at (-0.6,0.65) {$\mathrm{SM}_T$};
  \node at (-0.6,-0.15) {$\langle T_{\mu\nu}^{\Scal}T^{\Scal,\mu\nu}\rangle$};
  \draw[fill=gray, draw=gray] (0.75,0) circle (0.08cm);
  \draw[decorate, color=gray, decoration={snake, amplitude=1.5pt, segment length=7pt}] (0.25,0) -- (0.75,0);
  \draw[color=gray]        (0.75,0) -- (1.75, 0.70);
  \draw[dashed,color=gray] (0.75,0) -- (1.85, 0.35);
  \draw[dashed,color=gray] (0.75,0) -- (1.90, 0.00);
  \draw[dashed,color=gray] (0.75,0) -- (1.85,-0.35);
  \draw[color=gray]        (0.75,0) -- (1.75,-0.70);
  \node at (1.55,-1.0) {CFT};
  \node at (3,0) {$\Longleftrightarrow$};
\end{tikzpicture}
\begin{tikzpicture}[line width=1.3pt, scale=1]
  \path[use as bounding box] (-2.6,-2.2) rectangle (2.6,2.2);
  \draw[fill=gray!12, draw=black] (-2.05,0.75) circle (0.18cm);
  \draw[fill=gray!12, draw=black] (-2.05,-0.75) circle (0.18cm);
  \node at (-2.05,1.2) {$\mathrm{SM}_T$};
  \node at (-2.05,-1.2) {$\mathrm{SM}_T$};
  \draw[] (-1.87,0.75) -- (-0.45,0.15);
  \draw[] (-1.87,-0.75) -- (-0.45,-0.15);
  \draw[fill=gray, draw=gray] (-0.45,0) circle (0.12cm);
  \draw[decorate, decoration={snake, amplitude=1.5pt, segment length=7pt}]
    (-0.45,0) -- (0.65,0);
  \node at (0.10,0.42) {$\gamma_{\mu\nu}$};
  \draw[fill=gray, draw=gray] (0.65,0) circle (0.10cm);
  \draw[color=gray]        (0.65,0) -- (2.05, 0.80);
  \draw[dashed,color=gray] (0.65,0) -- (2.05, 0.40);
  \draw[dashed,color=gray] (0.65,0) -- (2.05, 0.00);
  \draw[dashed,color=gray] (0.65,0) -- (2.05,-0.40);
  \draw[color=gray]        (0.65,0) -- (2.05,-0.80);
  \node at (2.15,0) {CFT};
\end{tikzpicture}
\caption{\label{fig:GFI}
Equivalent representations of thermal gravitational freeze-in into the CFT sector that contains the dark matter.}
\end{figure}

This leads to the inclusive production of the DM as 
\be\label{eq:NumeroGFI}
N_{\rm SM \to DM} =-\frac{2c_J}{7680\Mpl^2}   \int \frac{d^4q}{(2\pi)^4} \langle T_{\mu\nu}^{\Scal}(q) T_{\rho\sigma}^{\Scal}(-q)\rangle \frac{\Pi^{\mu\nu\rho\sigma}(q)}{(q^2)^2}\theta(q^2)\,. 
\ee
We just need to compute the thermal average of $T^{\Scal}_{\mu\nu}T^{\Scal\,,\mu\nu}$, therefore we need to derive
\be
\langle T_{\mu\nu}(q)T_{\rho\sigma}(-q)\rangle_T \equiv \int d^4x d^4y e^{+i q (x-y)}\mathrm{tr}[\rho_T  T_{\mu\nu}(x)T_{\rho\sigma}(y)]= V_4 \int d^4 x e^{i q x}\mathrm{tr}[\rho_T  T_{\mu\nu}(x)T_{\rho\sigma}(0)]
\ee
where $\rho_T=e^{-H/T}$ is a thermal state at equilibrium temperature $T$, and in the second step we have exploited that translational invariance is preserved at finite temperature. Here $V_4$ is a four-dimensional volume factor. As self-evident, the calculation requires the knowledge of the two-point function of the energy momentum tensor at finite temperature in real space, $\mathrm{tr}[\rho_T  T_{\mu\nu}(x)T_{\rho\sigma}(0)]$. It can be computed taking the trace over a complete set of energy eigen-states. Also, by including a complete set of (momentum) eigen-states $|P\rangle$, including vacuum, single-particle and multi-particle states etc., we arrive at the following formal expression
\be
\mathrm{tr}[\rho_T  T_{\mu\nu}(x)T_{\rho\sigma}(0)] = \int \frac{d^4p}{(2\pi)^4}\frac{d^4P'}{(2\pi)^4}e^{-E_p/T} \langle p|T_{\mu\nu}(x)|P'\rangle \langle P'|T_{\rho\sigma}(0)|p\rangle\,.
\ee
By means of translational invariance $\langle p|T_{\mu\nu}(x)|P'\rangle=e^{i (p-P')x}\langle p|T_{\mu\nu}(0)|P'\rangle$. This, together with the fact that from the completeness relation we just retain the vacuum contribution, allows us to eventually land on the following expression
\be
\langle T_{\mu\nu}(q)T_{\rho\sigma}(-q)\rangle_T = V_4 e^{-q_0/T} \langle q|T_{\mu\nu}|0\rangle \langle 0|T_{\rho\sigma}|q\rangle + \cdots = V_4 e^{-q_0/T} \, 2\im[i\langle T T_{\mu\nu}(q) T_{\rho\sigma}(-q)\rangle] +\cdots
\ee
By neglecting the corrections represented by the dots,\footnote{These terms are associated the quantum statistics. In \cite{Redi:2020ffc} it was found that they lead to correction of the abundance $O(10)\%$. 
A complete thermal computation will appear elsewhere.} we can just plug the above expression into eq.~\eqref{eq:NumeroGFI} and read out the particle number. We notice that $N_{\rm dec}$ is extensive, being proportional to $V_4$, so that it makes sense to compute the production rate per unit time and volume, $\gamma\equiv N_{\rm SM\to DM}/V_4$, as follows
\be
\gamma_{\rm SM \to DM}=-\frac {2c_J}{7680 \pi} \frac 1{M_{\rm Pl}^4}  \int \frac{d^4q}{(2\pi)^4} \theta(q^2)  \frac 1 {(q^2)^2} e^{-q_0/T} 2\im[i\langle T T_{\mu\nu}(q) T_{\rho\sigma}(-q)\rangle] \Pi^{\mu\nu\rho\sigma}(q)\,.
\ee
We see that in this expression there is no footprint of the quantum statistics of the fields, which can be captured by retaining the full two-point function. In presence of a weakly coupled description this can be done as in \cite{Redi:2021ipn}. In the special case where the thermal state is a CFT with central charge $c_{\rm SM}$, as it is the case for the SM at very high temperatures, the production rate of the dark sector can be read out as
\begin{equation}
\begin{split}
\gamma_{\rm SM\to DS}&=  \frac 1{M_{\rm Pl}^4}\frac {2c_J}{7680 \pi}    \frac{2 c_{SM}}{7680\pi}\int  \frac{d^4q}{(2\pi)^4} \theta(q^2)e^{-\frac {q_0}T}\frac{\Pi^{\mu\nu\rho\sigma}(q) \Pi_{\mu\nu\rho\sigma}(q) }{q^4}\nonumber\\
&=\frac 1{M_{\rm Pl}^4} \frac {2c_J}{7680 \pi}   \frac{2 c_{SM}}{7680\pi} \int \frac{d^4q}{(2\pi)^4} \theta(q^2)e^{-\frac {q_0}T} 180 q^4\,.
\end{split}
\end{equation}
To perform the integral we introduce $s=\vec q\,^2$. Integration over $q_0$ gives,
\begin{equation}
\gamma_{\rm SM\to DS}=\frac {2}{M_{\rm Pl}^4}  \frac {2c_J}{7680 \pi}   \frac{2 c_{SM}}{7680\pi} \int ds \frac{s^{5/2} T K_1\left(\sqrt{s}/T\right)}{8 \pi ^3}=\frac{3 {c_J c_{SM}}}{1280 \pi ^5}\frac{T^8} {M_{\rm Pl}^4}\,.
\end{equation}
This agrees with the gravitational freeze-in computation \cite{Redi:2020ffc}, and demonstrates the equivalence of gravitational freeze-in 
with production from inhomogeneities.

As a cross-chek we can show that for a free thermal sector, the production of GWs is negligible at this level. The thermal two-point function is zero on a null-like support and the rate of graviton production vanishes. This is consistent with the findings of \cite{Ghiglieri:2015nfa,Ghiglieri:2020mhm}. Indeed to produce GW from the thermal SM we need to identify contributions to the anisotropic stress: they can arise from interactions on short distances or from the hydrodynamics limit at large scales. At weak coupling the largest contributions arises from two-loop corrections to the graviton propagator, which from the point of view of our computations correspond to adding a dissipative term: it is possible to emit gravitons on the time-like support of $\langle T^{\Scal}_{\mu\nu}(p)T^{\Scal}_{\rho\sigma}(-p)\rangle$ if they are produced in association with extra radiation \cite{Ghiglieri:2020mhm}. It would be interesting to extend the results of \cite{Ghiglieri:2020mhm} in a more general context.

\section{First order phase transitions}\label{sec:PT}

In this section we apply our results to first order phase transitions that are widely studied in the context of stochastic GW backgrounds.
This allows to explicitly check the relation between particle and GW production.

Cosmological first order phase transitions are violent events in the early universe that could have generated significant inhomogeneities
in the metric of spacetime. When the phase transition proceeds on a fast time scale $\beta^{-1}$, much faster than a Hubble time $\beta/H\gg 1$, the dynamics effectively develops in flat space. A first order phase transition is often described by a scalar field $\phi$ tunneling to its true vacuum from a metastable minimum gaining an energy $\Delta V(\phi)$. The time scale $\beta^{-1}$ of the transition is then related to the time variation of the bounce action in proximity of the nucleation time, when bubbles of true vacuum are created. The transition completes when bubbles collide and the universe is eventually in the true vacuum. 

Collision of bubbles, possibly affected by plasma, sound and turbulence effects corresponds to our source sector $\Scal$ in section \ref{sec:micro}. We are interested in the two-point function of the stress-energy tensor at bubble collisions $\langle T_{\mu\nu}(\vec q, \tau) T_{\rho\sigma}(-\vec q,\tau')\rangle$. This is known to generate a non-zero TT $\Pi_{\rm GW}$ power spectrum that sources both GWs and in our context also particle production. The presence of a GW background from bubble dynamics is one of the strongest motivation to study first order phase transitions, since they predict a unique spectrum with amplitude $\Omega_{\rm GW}$ and with a peak frequency observable in GW experiments. The abundance of GW is given by,
\begin{equation}
\Omega_{\rm GW}\equiv \frac 1 {\rho_c} \frac {d \rho_{\rm GW}}{d \log q} = \left(\frac{a_*}{a_0}\right)^4\frac{1}{\rho_{c,0}}\frac{1}{\Mpl^2} \frac{q^3}{2\pi^2} \Pi_{\rm GW}(q,q,-q)
\label{eq:GWfraction}
\end{equation}
The power spectrum today is peaked at frequency $f_{\rm peak}$ that is directly related to the temperature or scale factor of the universe when the background was 
produced. At production the peak frequency is of order $\beta$. Assuming that the phase transition occurs during radiation domination one finds,
\begin{equation}
f_{\rm peak}\sim 10^{-4}\,{\rm mHz} \frac {\beta}{H_*}\frac {T_*}{100\, {\rm GeV}}\,.
\label{eq:frad}
\end{equation}
The crucial parameter that characterizes the phase transition is its duration, typically of order $\beta^{-1}$.
In particular the amplitude of GW produced is suppressed for large $\beta$ while the frequency is enhanced. 
The other important parameter is velocity of the walls $v_w$ that can be subluminal or reach relativistic speed depending 
on the interactions with plasma of other particles. As we mentioned, the GW signal has several contributions that depend on the details of the phase transition and have different dependence on the parameters. However, here we focus on the collision of bubbles that can lead to strong gravity wave production. 

\begin{figure}[t]
\begin{center}

\includegraphics[width=0.45\textwidth]{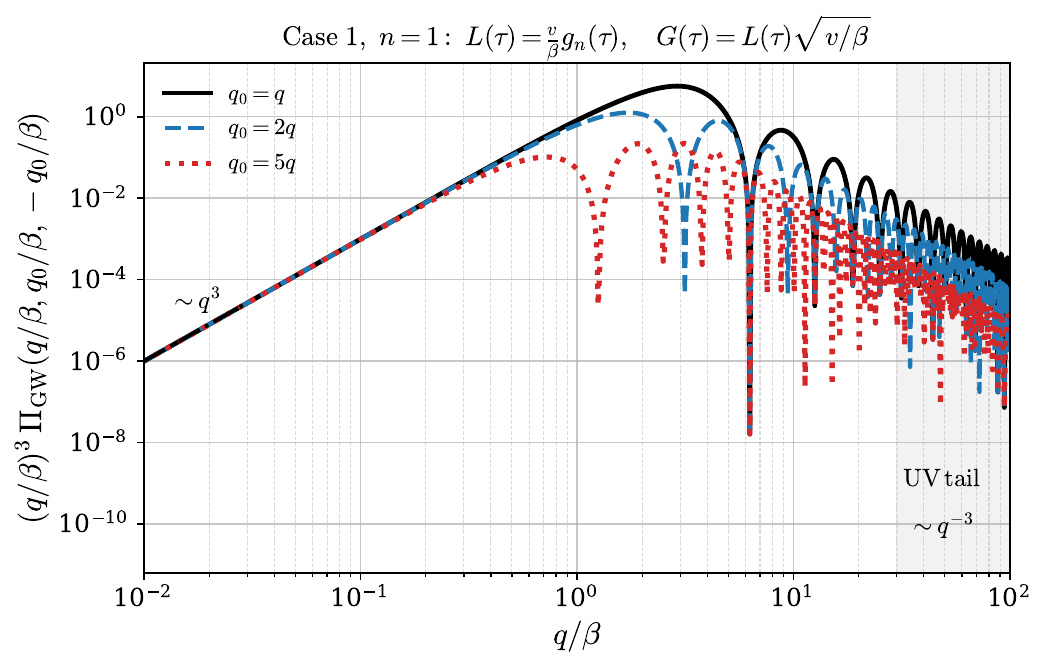}
\quad
\includegraphics[width=0.45\textwidth]{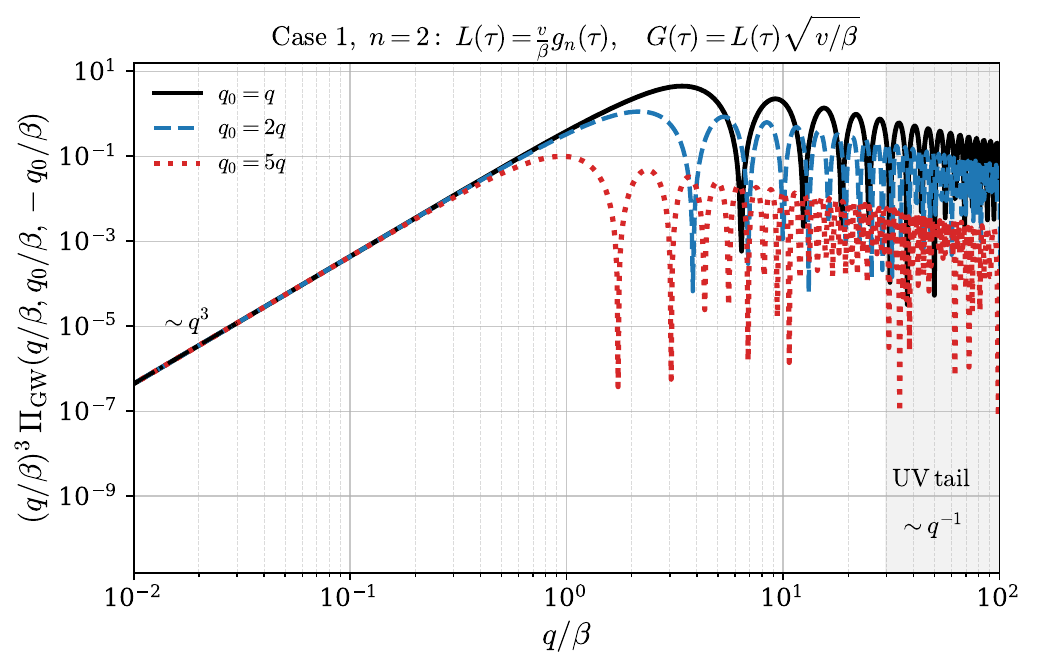}

\vspace{0.4cm}

\includegraphics[width=0.45\textwidth]{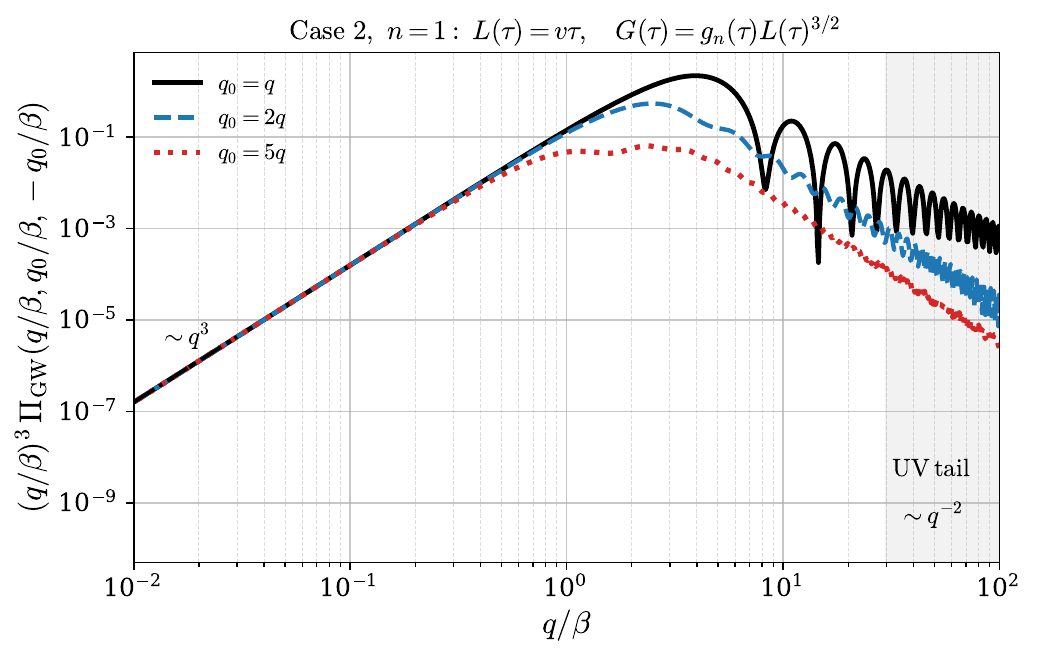}
\quad
\includegraphics[width=0.45\textwidth]{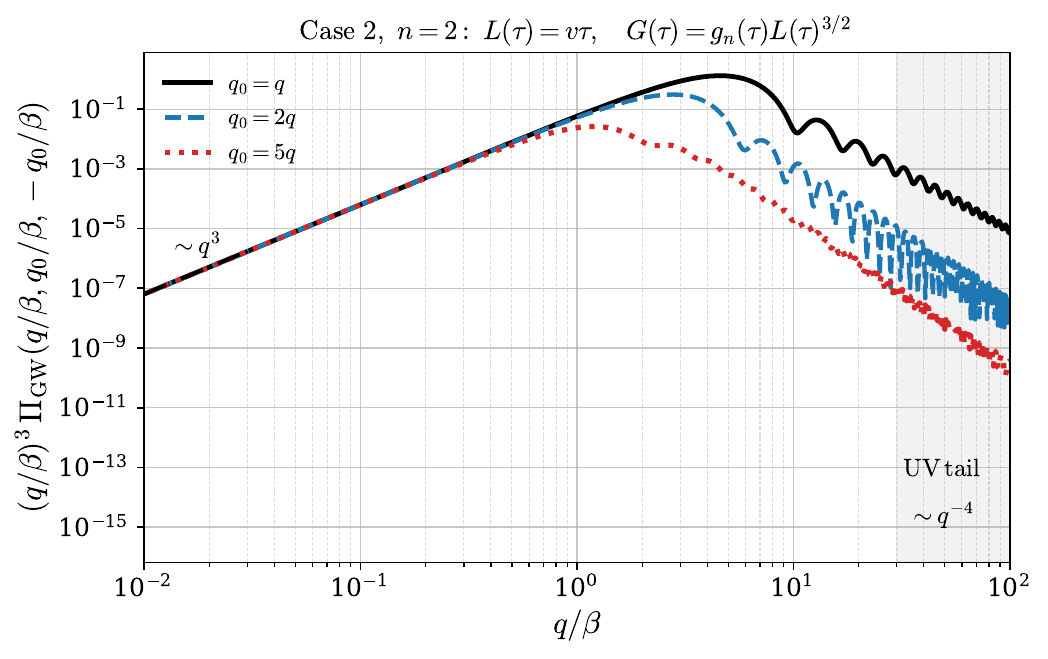}
\caption{
\textit{
Equal frequency power spectrum for first order phase transitions.
Top panels correspond to template 1 and bottom panels to template 2.
Blue, red, and dashed red curves refer to $q_0/q = 1,2,5$.
Left panels use the discontinuous function $g_1(\tau)$ while right panels use the $C^0$ function $g_2(\tau)$ from eq.~(\ref{eq:gn}).
}}
\label{fig:PTcombined}
\end{center}
\end{figure}

As discussed in the previous sections the knowledge of the power spectrum $\Pi_{\rm GW}$ in its full phase space, not only allows to compute gravity waves \eqref{eq:GWfraction}
but also particle production. The power spectrum $\Pi_{\rm GW}(q,\tau,\tau')$, can be determined through simulations or  analytically using  approximations 
such as the envelope approximation. Here will use the results in \cite{Caprini:2009fx,Caprini:2007xq} where several templates for $\Pi_{\rm GW}$ were studied that
capture the main features obtained in simulations. The purpose of this analysis is to show that the GW spectrum controls the
abundance of spectator particles produced. We here focus on a coherent source, where the unequal time two-point function factorizes
\begin{equation}
\Pi_{\rm GW}(q,\tau,\tau')\equiv f(q\,,\tau)f(q\,,\tau')\,.
\end{equation}
Using the appropriate templates for $f(q,\tau)$ we can directly compute the double Fourier transform in time, and use $\Pi_{\rm GW}(q,q_0,-q_0)$ to compute GW and particle production using eqs. (\ref{eq:abGWPP}). We expect $f(q,\tau)$ to have a time support comparable to the duration of the phase transition $\sim \beta^{-1}$, with a build up of the signal over such time scale that can be smooth or abrupt. In Ref. \cite{Caprini:2009fx} the following form for $f(q,\tau)$ was advocated
\begin{equation}
f(q\,,\tau)= G(\tau)\sqrt{ \frac {1+\left(\frac {kL(\tau)}3\right)^2}{1+\left(\frac {kL(\tau)}2\right)^2+\left(\frac {kL(\tau)}3\right)^6}} \,,
\label{eq:ansatzPT}
\end{equation}
where $G(\tau)$ and $L(\tau)$ are functions that grow on a time scale $\beta^{-1}$.
We consider two cases
\be
\begin{split}
\text{case 1)}& \quad G(\tau)=L(\tau) \sqrt{v/\beta}\,,\quad\quad L(\tau)=(v/\beta) g_n(\tau)\,, \\
\text{case 2)}& \quad G(\tau)=g_n(\tau) L(\tau)^{3/2}\,,\quad\quad L(\tau)=v\tau\,.
\end{split}
\ee
The functions $g_n(\tau)$ have support $\tau\in[0,\beta^{-1}]$ and are defined by 
\begin{equation}
g_n(\tau)=[4 \beta ^2 \tau(1/\beta-\tau)]^{n-1}\theta (\tau) \theta (1/\beta-\theta)\,.
\label{eq:gn}
\end{equation}
Notice that the case $n=1$ correspond to a discontinuos $f(q,\tau)$ that only last $\beta^{-1}$, while for $n>1$ the function describes 
 a signal that is built on a time scale $1/\beta$. 

With these parametrization we can determine numerically the power spectrum in Fourier space.  
In Figs. \ref{fig:PTcombined}  we show the power spectrum obtained from (\ref{eq:ansatzPT} ) for different values of $q_0/q$. 
The black curve corresponds to $q_0=q$ that determines the production of GW. 
On the other hand $q_0>q$ is what determines particle production. 
The abundance remains finite even for a discontinuous transition. Indeed for the templates in eq.~\eqref{eq:ansatzPT},
$f(q\,,\tau)$ goes to zero for large $q$ as $1/q^2$. Given that the time Fourier transform goes as $1/q_0$ it follows that the total abundance is finite.

The analysis was carried out for GW where particle production is directly related to production of GW.
In phase transition scalar perturbations are also produced.  On general grounds we do not expect a large difference between scalar and tensor perturbations
but a detailed analysis would be needed. In fact production of tensor modes requires the existence of anisotropic stress to produce a quadrupole. This can be suppressed
by interactions, i.e. a perfect fluid has no anisotropic stress. As a consequence scalar perturbations could be more important 
in realistic scenario of phase transitions. This question could be studied with existing numerical simulations.

Let us briefly discuss incoherent backgrounds.
In \cite{Caprini:2009fx} it was considered the possibility that the two-point function of the anisotropic stress is fully incoherent
so that it is proportional to $\delta(\tau-\tau')$ that leads to consistent results for gravity wave production. 
In this case however $\Pi(q\,,q_0\,,-q_0)$ does not depend on $q_0$ so that the total number of particles is divergent.
A fully incoherent background is unphysical as it corresponds to at a singular spacetime. 
It is however possible that incoherent backgrounds enhance the signal.

\section{Conclusions}\label{sec:conclusions}

This work together with \cite{Garani:2025qnm} lays out the general formalism to compute the abundance of a dark sector 
that is only coupled to the SM gravitationally. This encompasses in a simple framework gravitational production 
from the SM thermal bath as well as recently discussed particle production from inhomogeneities.  
One general result emerging is the equivalence between particle production as propagation in inhomogeneous background 
and gravitational scattering. 

In general particles can be produced due to cosmological evolution of perturbations or local non adiabatic processes such as first order phase transitions or preheating.
We have shown that for sources well inside the Hubble horizon the result can be related to the expectation value of the two-point function of the energy momentum tensor that creates the perturbations. 
Moreover for conformally coupled sectors the abundance is proportional to the central charge of the sector $c_J$. This provides an unavoidable mechanism to produce
DM of a dark sector that can dominate compared to other contributions discussed in the literature.

Remarkably for tensor perturbations  this mechanism is closely related to the production of GW being determined
by the very same two-point function of the transverse traceless components of the energy momentum tensor, albeit in a different kinematical regime.  
This allows to relate the abundance of the dark sector to the energy fraction and frequency of GW today. 
In order to obtain cosmologically significant abundances the frequency must be very large, corresponding to production at early times. 
Indeed the numerical abundance scales on general grounds  as $f^3$ making it negligible in the region for $f< 100$ Hz, 
relevant for current gravitational wave experiments.

Scalar and vector perturbations of the metric lead to particle production while they do not source GW.
Such perturbations are expected to be produced along with tensor perturbation in realistic scenarios that lead to significant inhomogeneities.
Our estimates indicate that the abundance is comparable or larger to tensor perturbations but scalar and vector perturbation
are poorly studied in the literature since they do not produce GW. This motivates the study of these modes in explicit examples 
such as  first order phase transitions or preheating. We are planning to study these effects with numerical and analytical methods in future work \cite{future}.
In the context of first order phase transitions, it is also interesting to study the two-point function of the transverse-traceless part of the energy-momentum tensor in the full phase space. 
This information could, in principle, be extracted from existing simulations.

For production from a thermal bath the result  is controlled by the thermal two-point function of the energy momentum
tensor in real time for which powerful methods have been developed in the context of GW production, see \cite{Ghiglieri:2015nfa,Ghiglieri:2020mhm}. 
We are planning to extend these results to particle production allowing to capture the effect of interactions and quantum effects.

\appendix

\subsubsection*{Acknowledgements}
{\small
We wish to thank Stefano Bartolomei and  Azadeh Maleknejad for discussions.
We gratefully acknowledge the Galileo Galilei Institute for Theoretical Physics (Florence) for hospitality. 
This work is supported by the Italian Ministry of University and Research (MUR) via the PRIN 2022 project n. 20228WHTYC (CUP:I53C24002320006).
}

\appendix

\section{Metric perturbations in FRW}
In this appendix we derive the full unequal time correlators for scalar $\Theta$, vector $V$ and tensor $h$ perturbations in the FLRW background, generalize our derivation in flat space of section \ref{sec:micro}. In FLRW, the simplicity of the connection between the sources $\Pi_{\Theta, V, \rm GW}$ and the power spectra $\Delta_{\Theta,V,\rm GW}$ is lost due to the appearance of Hubble and scale factors in the Einsteins' equations of eq.~\eqref{eq:MB}.

The approach here is that, once we neglect the primordial fluctuations generated by inflation, the solutions for the perturbations are just the convolution of the retarded Green's functions $G_{\rm R}^{\Theta,W,h}(\vec q; \tau,\tau')$ and the sources $S_{\Theta,W,h}(\vec q, \tau')$, which may differ from the ones in flat space. The retarded Green's functions instead are solutions form the cosmological perturbations in FLRW in response to a Dirac delta function in cosmological time per each Fourier mode for positive time difference. We aim at \begin{eqnarray}
\Theta(\vec q, \tau)&=&\int d\tau' G^{\rm R}_{\Theta}(\tau,\tau'; \vec q) S_\Theta(\vec q, \tau')\,,\quad \Theta(\vec q, q_0)=\int \frac{dq_0'}{(2\pi)}G^{\rm R}_{\Theta}(q_0,q_0';\vec q) S_\Theta(\vec q, -q_0')\\
W(\vec q, \tau)&=&\int d\tau' G^{\rm R}_{W}(\tau,\tau'; \vec q) S_W(\vec q, \tau')\,,\quad V(\vec q, q_0)=\int \frac{dq_0'}{(2\pi)}G^{\rm R}_{W}(q_0,q_0';\vec q) S_W(\vec q, -q_0')\,,\\
h(\vec q, \tau)&=&\int d\tau' G^{\rm R}_{h}(\tau,\tau'; \vec q) S_h(\vec q, \tau')\,,\quad h(\vec q, q_0)=\int \frac{dq_0'}{(2\pi)}G^{\rm R}_{h}(q_0,q_0';\vec q) S_h(\vec q, -q_0')\,.
\end{eqnarray}
In FLRW, due to the Hubble friction, the Green's function is not conformal time-translational invariant, therefore $G^{\rm R}$ is not proportional to a $\delta(q_0-q_0')$.
By manipulation of the linearized Einsteins' equations, combining eq.s [\eqref{eq:drho} - $c_a^2$ \eqref{eq:dp}] using the constraint in \eqref{eq:sigma}, we have
\begin{eqnarray}
\Theta''(\vec q, \tau) + 3(1+c_a^2)\mathcal H\,\Theta'(\vec q, \tau)  + \left[ c_a^2 q^2 + 2\mathcal H' + (1+3c_a^2)\mathcal H^2\right]\Theta(\vec q, \tau)    &=&   S_\Theta(\vec q, \tau)\,,\\
-q^2 W_i^{T\,\prime} &=&S_W\\
h_{ij}''+ 2 \mathcal{H}h_{ij}' + q^2 h_{ij} &=&S_{ij}^h
\end{eqnarray}
The retarded Green's functions are solutions to this equation where the source is $\delta(\tau-\tau')$ for $\tau>\tau'$. Notice that here $\mathcal{H}$ and the adiabatic sound speed $c_a^2\equiv \bar{p}'/\bar{\rho}'$ depends on all the contributions (including the source). The explict expressions for the sources are the following 
\begin{eqnarray}
S_\Theta &\equiv& \frac{a^2 (\delta p - c_a^2 \delta \rho)}{M_{\rm Pl}^2} + \left[ \partial_\tau^2 + (1+3c_a^2)\mathcal H\,\partial_\tau + \left[(c_a^2+\frac23)q^2 - 2\mathcal H'
- (1+3c_a^2)\mathcal H^2\right] \right] \big(\frac{3a^2 \hat{\sigma}}{2M_{\rm Pl}^2q^2}\big)\,\\
S_W &\equiv& \frac{2}{M_{\rm Pl}^2}\,
a^2\left(\bar\rho+\bar p\right)\delta T_i^{0\,{\rm T}}\,,\\
S_h &\equiv&\frac{2}{\Mpl^2}T_{ij}^{\rm TT}\,.
\end{eqnarray}
where we have defined $\hat\sigma\equiv (\rho+p)\sigma$.
Here $S_\Theta$ is generic, we can however assume that only the source sector has non-negligible non-adiabatic components and anisotropic stress. We can check that we get the correct flat space limit by enforcing $a\to 1$, $\mathcal H\to 0$. The solution in Fourier transform is
\be
\Theta(\vec q,q_0)\stackrel{a\to 1,\mathcal{H}\to 0}{=}\frac{\left[\delta p -c_a^2 \delta \rho + (1 + \frac{3}{2}c_a^2 - \frac{3}{2}\frac{q_0^2}{q^2})\hat\sigma\right]}{(-q_0^2+c_a^2 q^2)\Mpl^2} \stackrel{ \partial_\mu T^{\mu\nu} =0}{=}\frac{1}{q^2\Mpl^2}(\delta\rho -\frac32\hat\sigma)
\ee
where in the second step we have used flat-space stress-energy tensor conservation, $\delta p = q_0^2/q^2 \delta\rho -\hat\sigma$. We see that the apparent acoustic pole at $q_0=c_a |q|$ is unphysical and we recover the Poisson equation of \eqref{eq:TTheta}, once the conservation of $T_{\mu\nu}$ is considered.

\subsection{Scalar fluctuations: single fluid dominance}
In the limit where the source is a small perturbation to the background geometry, we can write $c_a^2=w=\mathrm{constant}$ and $2\mathcal H' + (1+3c_a^2)\mathcal H^2=0$ if only one barotropic component is dominating the universe. For scalar perturbations the left-hand side becomes $\Theta'' +3(1+w)\mathcal{H}\Theta' + w q^2 \Theta$. This allows us to construct the Green's function more easily for radiation or matter dominance. Also the source term $S_\Theta$ simplifies in this limit, as it depends on background quantities.

\begin{itemize}
\item \textbf{Radiation dominance.} Here $w=c_s^2=1/3$ and $\mathcal{H}=1/\tau$. The retarded Green's function is
\be
G_{\Theta}^{\rm R}\big|_{\rm rad.}=\theta(\tau-\tau')\,
\frac{\tau'}{\left(q^2/3\right)^{3/2}\tau^3}
\left[
\left(1+\frac{q^2}{3}\tau\tau'\right)
\sin\!\left(\frac{q}{\sqrt{3}}(\tau-\tau')\right)
-
\frac{q}{\sqrt{3}}(\tau-\tau')
\cos\!\left(\frac{q}{\sqrt{3}}(\tau-\tau')\right)
\right]
\ee
While the expression of the source simplifies to
\be
S_\Theta|_{\rm rad.}=\frac{a^2}{\Mpl^2}(\delta p - \frac13 \delta \rho) + [\partial_\tau^2 +2\mathcal H \partial_\tau  +q^2]\left(\frac{3a^2}{2M_{\rm Pl}^2q^2}(\rho+p)\sigma\right)
\ee
\item \textbf{Matter dominance.} Here $w=0$ and $\mathcal{H}=2/\tau$.
The retarded Green's function is
\be
G_{\Theta}^{\rm R}\big|_{\rm mat.}=\theta(\tau-\tau') \frac{\tau'}{5}
\left[
1-\left(\frac{\tau'}{\tau}\right)^5
\right]
\ee
While the expression of the source simplifies to
\be
S_\Theta|_{\rm mat.}=\frac{a^2}{\Mpl^2}\delta p + [\partial_\tau^2 +\mathcal H \partial_\tau +\frac23 q^2]\left(\frac{3a^2}{2M_{\rm Pl}^2q^2}(\rho+p)\sigma\right)
\ee
\end{itemize}

\subsection{Tensor Perturbations}

Using the rescaled field $H_{ij}=a h_{ij}$ the equation (72) becomes,
\begin{equation}
H_{ij}''+\left(q^2-\frac{a''}{a^2}\right)H_{ij}= \frac 2 {M_{\rm Pl}^2} a T_{ij}^{TT}
\end{equation}
so that,
\begin{equation}
h_{ij}(q\,,\tau)= \frac 2 {M_{\rm Pl}^2 a(\tau)} \int_{-\infty}^{\infty} d\tau' G_q(\tau-\tau') a(\tau')T_{ij}^{TT}(q\,,\tau')
\end{equation}
where $G_q(\tau-\tau')$ is in general the Green's function with retarded boundary conditions.
In radiation domination $a=a_* \tau$ so the left hand side is just a free harmonic oscillator.
The Green's function is then,
\begin{equation}
G_q^{\rm rad}(\tau)= \theta(\tau) \frac{ \sin q \tau}{2q}
\end{equation}
This allows to write,
\begin{equation}
\Delta_h(q\,,\tau\,,\tau')= \frac 4 {M_{\rm Pl}^4 a_*^2 \tau \tau'}\int_{-\infty}^{\infty} dt dt' \theta(\tau-t) \frac{ \sin q (\tau-t)}{2q}\theta(\tau'-t') \frac{ \sin q (\tau'-t')}{2q} \Pi_{GW}(q\,,t\,,t')
\end{equation}
This equation is on the same footing of the general expression for the production of stochastic GW backgrounds, see \cite{Caprini:2018mtu}
The Fourier transform determines exactly particle production including the cosmological expansion.

\pagestyle{plain}
\bibliographystyle{jhep}
\small
\bibliography{biblio_stochastic_sources}

\end{document}